\documentclass[final,1p,times]{elsarticle}

\makeatletter
\def\ps@pprintTitle{%
   \let\@oddhead\@empty
   \let\@evenhead\@empty
   \def\@oddfoot{\reset@font\hfil\thepage\hfil}
   \let\@evenfoot\@oddfoot
}
\makeatother

\biboptions{sort&compress,square}
\usepackage{empheq}
\usepackage{todonotes}
\usepackage{bbm}
\usepackage{siunitx}
\usepackage{upgreek}
\usepackage{booktabs}
\usepackage[hyphens]{url}
\usepackage{hyperref}
\hypersetup{breaklinks=true, hidelinks}
\newcommand {\gz}[1]{\boldsymbol{#1}}
\newcommand{\eqn}[1]{Eq.~(\ref{#1})}

\newdefinition{rmk}{Remark}

\def\Grad {{\rm Grad}}
\def\grad {{\rm grad}}
\def\Div {\mbox{Div\,}}

\def\d {\,\mbox{d}}
\def\D {\,\mbox{D}}
\def\tdot   {\;\cdot\!\!:}

\def \motion {\gz \varphi}
\def \F {\gz F}
\def \invF {\gz f}
\def \mF {\overline{\gz F}}
\def \invmF {\overline{\gz f}}
\def \mG {\overline {\gz G}}
\def \potential {\varphi}
\def \Efield {\mathbb{E}}
\def \efield {\mathbbm{e}}
\def \x {\gz x}
\def \X {\gz X}
\def \b {\gz b}
\def \Ptot {\gz P^{\text{tot}}}
\def \P {\gz P}
\def \Ppol {\gz P^{\text{pol}}}
\def \Pmax {\gz P^{\text{max}}}

\def \mP {\overline {\gz P}}
\def \mQ {\overline {\gz Q}}
\def \eD {\mathbb{D}}
\def \eP {\mathbb{P}}

\def \rhob {\rho_0^f}
\def \rhos {\widehat{\rho}_0^f}

\def \mt {\overline{\gz t}_0}
\def \t {\gz t_0}
\def \smF {\overline{\mathbf{\mathsf{F}}}}
\def \smotion {\gz \upvarphi}
\def \spotential {\upvarphi}

\newcommand{\fig}[1]{Fig.~\ref{#1}}

\def\mcl  #1{               {\cal #1}}

\newcommand{\trns}{{}^{\mathsf{T}}}
\newcommand{\invtrns}{{}^{-\mathsf{T}}}
\newcommand{\cof}{\text{cof}}

\newcommand{\sect}[1]{Sec.~\ref{#1}}

\title{Modelling the flexoelectric effect in solids: a micromorphic approach}
\author[label1]{A.T. McBride}
\author[label2]{D. Davydov}
\author[label1,label2]{P. Steinmann}
\address[label1]{Glasgow Computational Engineering Centre, University of Glasgow, United Kingdom}
\address[label2]{Chair of Applied Mechanics, Friedrich-Alexander University of Erlangen--Nuremberg, Germany}

\begin{document}

\begin{frontmatter}

\begin{abstract}
    Flexoelectricity is characterised by the coupling of the second gradient of the motion  and the electrical field in a dielectric material.
    The presence of the second gradient is a significant obstacle to obtaining the approximate solution using conventional numerical methods, such as the finite element method, that typically require a $C^1$-continuous approximation of the motion.
    A novel micromorphic approach is presented to accommodate the resulting  higher-order gradient contributions arising in this highly-nonlinear and coupled problem within a classical finite element setting.
    Our formulation accounts for all material and geometric nonlinearities, as well as the coupling between the mechanical, electrical and micromorphic fields.
    The highly-nonlinear system of governing equations are derived using the Dirichlet principle and approximately solved using the finite element method.
    A series of numerical examples serve to elucidate the theory and to provide insight into this intriguing effect that underpins or influences many important scientific and technical applications.
\end{abstract}

\begin{keyword}
    Flexoelectricity \sep Micromorphic Continua \sep Finite Element Method
\end{keyword}

\end{frontmatter}

\section{Introduction}

In a piezoelectric material, an applied \emph{uniform} strain can induce electric polarisation (or vice versa).
Crystallographic considerations restrict this important property to non-centrosymmetric systems.
By contrast, flexoelectricity\footnote{\emph{Flexo} and \emph{Piezo} derive from the latin words \emph{flecto} - to bend - and \emph{piezein} - to squeeze.  } is the property of an insulator whereby it polarises when subjected to an \emph{inhomogeneous deformation} (i.e.\ a strain gradient).
The inhomogeneous deformation breaks the material's centrosymmetry, thereby allowing polarisation in non-piezoelectric materials.
Flexoelectricity can occur in materials of any symmetry, broadening the range of materials for use as actuators and sensors \citep{Jiang2013}.
Flexoelectricity is therefore of considerable interest to the engineering community and is the subject of extensive research.

Flexoelectricity in solids was first identified by \citeauthor{Mashkevich1957} \citep{Mashkevich1957, Tolpygo1963} and the theoretical foundations laid by \citet{Kogan1964}.
Recent rapid advancements in the miniaturisation of fabricated components has stimulated substantial experimental research into the flexoelectric effect \cite{Ma2001, Ma2002, Zubko2007} as gradient effects are more pronounced at smaller length scales.
Structures at small length scales can also exhibit a size-dependent mechanical response \citep[see e.g.][]{Stelmashenko1993, Fleck1994}.
Thus any representative model for flexoelectricity needs to account for both the coupling of the electrical response to a strain-gradient, and  size-dependent mechanical effects.
Reviews on flexoelectricity include \citep{Tagantsev1987, Tagantsev1991, Maranganti2006, Ma2010, Nguyen2013, Lee2012, Zubko2013, Krichen2016}.

The flexoelectric effect can be classified as direct or converse.
Direct is when a strain gradient induces polarisation; converse is when an electric field gradient induces a mechanical stress.
The direct flexoelectric effect will allow novel piezoelectric composites containing no piezoelectric elements to be developed \citep{Ma2010}.
Flexoelectricity is also responsible for electromechanical behaviour in hard crystalline materials and underpins core mechanoelectric transduction phenomena in biomaterials \citep{Nguyen2013}.

Classical continuum theories are unable to account for the size-dependent response exhibited by structures at small length scales.
Extended continuum models have been actively developed over the past three decades to remedy this deficiency.
 A significant proportion of extended models are members of either the gradient or micromorphic frameworks.
 Micromorphic continua are characterised by additional degrees of freedom at each continuum point \citep{Eringen1999, Mindlin1964, Toupin1964}.
 By contrast, gradient continua possess higher gradients of their primary fields \citep[see][and the references therein]{Forest2009}.

 The purely mechanical micromorphic theory has been extended to account for electromagnetic coupling by including the additional classical continuum electrodynamic contributions in the balance relations \citep[see e.g.][]{Eringen2003, Eringen2004}.
\citeauthor{Romeo2011} directly accounted for electromagnetic contributions at the microscale in a micromorphic framework by accounting for electric dipole and quadrupole densities.
This theory was extended to account for dielectric multipoles \citep{Romeo2015, Romeo2020} and thereby describe the piezoelectric and flexoelectric effect.

Numerical models that capture the key physics of flexoelectricity for arbitrary geometries in three dimensions are however limited.
This is particularly true for soft dielectric materials that can undergo significant deformation.
A central impediment to developing finite element models for flexoelectricity, or indeed gradient elasticity, is the requirement that the basis functions used to approximate the displacement field must be piecewise smooth and globally $C^1$-continuous.
This constraint arises as the partial differential equation governing the mechanical problem is of fourth-order.
By contrast, one only requires a standard $C^0$-continuous approximation for electro-elasticity.
$C^1$-continuous finite element approximations for complex geometries in three space dimensions are limited \citep{Gomez2008}.
Options include isogeometric analysis \citep{Hughes2005}, mixed formulations, discontinuous Galerkin approximations \citep{Engel2002}, the natural element method \citep{Sukumar1999} and other specialised element formulations, and meshless methods \citep{Askes2002}.
Many of these methods are not easily implemented within a conventional finite element library.

Many of the aforementioned methods to generate $C^1$-continuous finite element approximations have been used to model the problem of flexoelectricity.
\citet{Abdollahi2014} chose a meshless method.
The analysis was restricted to two dimensions and to the linearised theory.
They recently extended the formulation to three dimensions to provide new insight into the pyramid compression tests used to characterise the flexoelectric parameters.
Related works include \citep{Abdollahi2015b, Abdollahi2015c}.
\citet{Deng2014} developed a nonlinear theory for flexoelectricity in soft materials and biological membranes.
Numerical results were restricted to one space dimension.
They used a fourth-order approximation for the displacement field.
This however is not sufficient for a global $C^1$-continuous finite element approximation.
A mixed formulation based on theory of generalised (extended) continua was proposed by \citet{Mao2016}.
The mixed approach allowed the linearised gradient theory to be treated within a standard $C^0$-continuous finite element setting.
In a key contribution, \citet{Yvonnet2017} extended the nonlinear theory of electroelasticity \citep[see e.g.][and references therein]{Dorfmann2005, Pelteret2016, Vu2006} to account for the coupling between polarization and the gradient of the deformation gradient $\gz{G}$ - a third-order tensor.
A non-standard, $C^1$-continuous, Argyris-triangle-based finite element formulation was used.
This restricts the approach to relatively simple geometries and two dimensions.
In contrast to the majority of flexoelectricity models, the free space surrounding the continuum body was accounted for.

A major contribution of the work presented here is to model the scale-dependent effects that underpin flexoelectricity (the direct effect) using the micromorphic approach.
The formulation is not restrictive and can handle arbitrary geometries in three space dimensions.
We exploit the Dirichlet principle to uncover the relations governing the response of a (soft) dielectric material exhibiting flexoelectric effects.
Both geometric and materials nonlinearities are accounted for.
The framework is flexible and allows one to describe a range of related problems via an appropriate restriction of the constitutive parameters.
Several forms for the flexoelectric energy are proposed.

The highly-nonlinear system of governing equations are solved approximately using the finite element method.
A Newton--Raphson strategy is used to linearise the problem.
The framework is robust and exploits distributed parallelisation and automatic differentiation to improve the efficiency and to simplify the implementation, respectively.
Parallelisation helps offset the increased computational cost that arises in the micromorphic approach due to the need to approximate the micro-deformation field, a second-order tensor, in addition to the motion and the electric potential.
The finite element model is implemented with the open-source library deal.II \citep{Bangerth2007, dealII91}.

The structure of the presentation is as follows.
The theoretical background is presented in \sect{sec_theory}.
This includes the kinematics of the macroscopic, micromorphic  (microscopic) and electric problems.
The governing equations and boundary conditions are then derived using the Dirichlet principle.
Concrete forms for the constitutive relations are also given.
Details of the monolithic finite element formulation are provided in \sect{sec_fe_approximation}.
The theory is then elucidated via a series of numerical example problems in \sect{sect_numerical_examples}.
The presentation concludes with a summary and discussion.

\section*{Notation}

Direct notation is adopted throughout.
Occasional use is made of index notation, the summation convention for repeated indices being implied.
Indices associated with the reference configuration and the current configuration of the body are distinguished by the use of upper- and  lower-case font, respectively.

The scalar products of two vectors $\gz a$ and $\gz b$, two second-order tensors  $\gz A$ and $\gz B$, and two third-order tensors $\gz Q$ and $\gz G$ are respectively denoted by
\begin{align*}
    \gz a \cdot \gz b = a_i b_i \, ,
    &&
    \gz A : \gz B = A_{ij} B_{ij} \, ,
    &&
    \gz Q \tdot \gz G := Q_{ijk} G_{ijk} \, .
\end{align*}
The conventional dyadic product of two vectors, and of two second-order tensors are respectively given by
\begin{align*}
    \gz a \otimes \gz b = a_i b_j \gz e_i \otimes \gz e_j
    && \text{and} &&
    \gz A \otimes \gz B = A_{ij} B_{kl} \gz e_i \otimes \gz e_j \otimes \gz e_k \otimes \gz e_l \, ,
\end{align*}
where $\gz e_i \in \mathbb{R}^{n^\text{dim}}$ and $\gz E_I \in \mathbb{R}^{n^\text{dim}}$ are the basis vectors of the Cartesian coordinate frame in the current (spatial) and reference (material) settings, respectively, and $n^\text{dim}$ is the space dimension.
The upper and lower dyadic products of pairs of second-order tensors are respectively given by
\begin{align*}
    \gz A \overline{\otimes} \gz B = A_{ik} B_{jl} \gz e_i \otimes \gz e_j \otimes \gz e_k \otimes \gz e_l
    &&
    \text{and}
    &&
    \gz A \underline{\otimes} \gz B = A_{il} B_{jk} \gz e_i \otimes \gz e_j \otimes \gz e_k \otimes \gz e_l  \, .
\end{align*}
The second-order identity tensor is defined by
\begin{align*}
    \gz I = \delta_{ij} \gz e_i \otimes \gz e_j \, .
    % &&
    % \mathbb{I} = \delta_{ik} \delta_{jl} \gz e_i \otimes \gz e_j  \otimes\gz e_k \otimes \gz e_l \, ,
    % &&
    % \widetilde{\mathbb{I}} = \delta_{il} \delta_{jk} \gz e_i \otimes \gz e_j \otimes \gz e_k \otimes \gz e_l
    %\, .
\end{align*}
The action of a second-order tensor $\gz A$ on a vector $\gz b$ is the vector $\gz c$ defined by
\begin{align*}
   \gz c = \gz A \cdot \gz b = A_{im} b_{m} \gz e_i \, .
\end{align*}
The single contraction of two second-order tensors, $\gz A$ and $\gz B$, is the second-order tensor $\gz C$ defined by
\begin{align*}
    \gz C = \gz A \cdot \gz B = A_{im} A_{mj} \gz e_i \otimes \gz e_j \, .
 \end{align*}

Micromorphic variables are distinguished from macroscopic quantities by an overline.
Variables associated with the electrical problem are distinguished using blackboard bold.
Further notation is introduced when required.

\section{Theoretical background}\label{sec_theory}

The kinematics of the coupled problem of flexoelectricity are presented in \sect{sect_kinematics}.
The Dirichlet principle is then employed to derive the governing equations and boundary conditions.
Concrete forms for the constitutive relations are then provided.

\subsection{Kinematics} \label{sect_kinematics}

The kinematic description of motion at the macroscopic scale is presented in \sect{sect_kinematics_macro}.
This is followed by the description of the micromorphic problem at the microscopic scale.
The electric problem is then given.
For more details on the formulation of coupled nonlinear electro-elasticity, see \citep{Dorfmann2005, Dorfmann2014, Steinmann2011} and for nonlinear micromorphic elasticity see \citep{Hirschberger2007}, and the references therein.

\subsubsection{The macroscopic problem}\label{sect_kinematics_macro}

\begin{figure}[htb!]
    \centering
    \includegraphics[width=\textwidth]{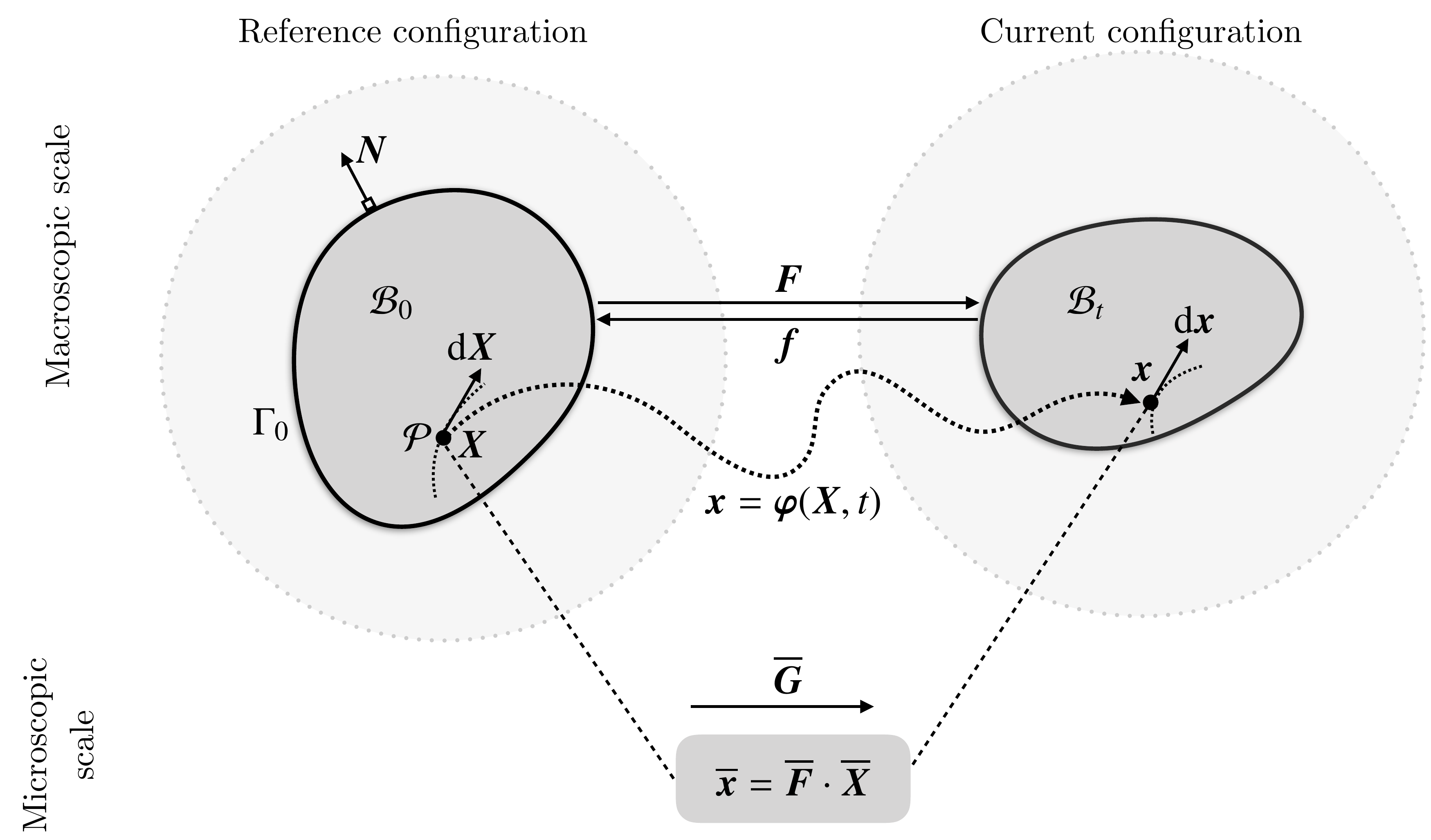}
    \caption{The reference and current configurations of the continuum body $\mcl B$ and the associated macroscopic and microscopic (micromorphic) motions and deformation gradients.}
    \label{fig_kinematics}
  \end{figure}

Consider a continuum body $\mcl B$ composed of matter as shown in \fig{fig_kinematics}.
The motion of $\mcl B$ from its reference configuration $\mcl B_0$ to its current configuration $\mcl B_t$
is defined via the map $\x = \motion (\X,t)$, where $\x \in \mcl{B}_t$ and $\X \in \mcl{B}_0$ are physical points in the current and reference configurations, respectively.
The boundary of the reference configuration is denoted by $\Gamma_0$, with outward unit normal $\gz{N}$.

The invertible linear tangent map $\F$ (i.e.\ the deformation gradient) maps a line element $\d \X$ in the reference configuration to a line element $\d \x$  in the current configuration and is defined by the derivative of the motion with respect to the material placement; that is,
\begin{align*}
    \F := \Grad \motion
    =
    \dfrac{\partial \varphi_i}{\partial X_J} \gz{e}_i \otimes \gz{E}_J
    && \text{and} &&
    \d \x = \F \cdot \d \X \, .
\end{align*}
The determinant of $\gz{F}$ is defined by $J := \det \gz{F} > 0$ and its inverse by $j := 1 / J$.
The symmetric right and left Cauchy--Green tensors, $\gz{C}$ and $\gz{b}$, are respectively defined by
\begin{align*}
    \gz{C} := \gz{F}\trns \cdot \gz{F} && \text{and} && \gz{b} := \gz{F} \cdot \gz{F} \trns \, .
\end{align*}
It proves convenient to define the inverse of the deformation gradient by $\gz{f} := \gz{F}^{-1}$.
The Piola strain $\gz{B}$ and the Finger strain $\gz{c}$ are respectively defined by
\begin{align*}
    \gz{B} := \gz{f}  \cdot \gz{f}\trns = \gz{C}^{-1}
    && \text{and} &&
    \gz{c} := \gz{f} \trns \cdot \gz{f} = \gz{b}^{-1} \, .
\end{align*}
Furthermore, the Green--Lagrange and Euler--Almansi strain tensors are respectively defined by
\begin{align*}
 \gz{E} := \dfrac{1}{2}\left[ \gz{C} - \gz{I} \right]
 && \text{and} &&
 \gz{e} := \dfrac{1}{2}\left[ \gz{i} - \gz{c} \right]
 \intertext{where}
 \gz{E} = \gz{F}\trns \cdot \gz{e} \cdot \gz{F} =: \gz{\varphi}^{-1}_\star (\gz{e})
 && \text{and} &&
 \gz{e} = \gz{f}\trns \cdot \gz{E} \cdot \gz{f} =: \gz{\varphi}_\star (\gz{E}) \, .
\end{align*}
The second-order identity tensors in the referential and current configurations are denoted by $\gz I$ and $\gz i$, respectively.
Note, $\invF \cdot \F = \gz I$  and $\F \cdot \invF = \gz i$.
The push-forward and pull-back operations on second-order tensors are denoted by $\gz{\varphi}_\star$ and $\gz{\varphi}^{-1}_\star$, respectively.
That is,
\begin{align*}
    \gz{\varphi}_\star (\bullet)^\flat =  \gz{f}\trns \cdot (\bullet)^\flat \cdot \gz{f}
    && \text{and} &&
    \gz{\varphi}^{-1}_\star (\bullet)^\flat =  \gz{F}\trns \cdot (\bullet)^\flat \cdot \gz{F} \, , \\
    \gz{\varphi}_\star (\bullet)^\natural =  \gz{F} \cdot (\bullet)^\natural \cdot \gz{F}\trns
    && \text{and} &&
    \gz{\varphi}^{-1}_\star (\bullet)^\natural =  \gz{f} \cdot (\bullet)^\natural \cdot \gz{f}\trns \, ,
\end{align*}
where $(\bullet)^\flat$ and $(\bullet)^\natural$ denote covariant and contravariant second-order tensors, respectively.
For completeness, the  push-forward and pull-back operations on vectors are given by
\begin{align*}
    \gz{\varphi}_\star (\bullet)^\flat =  \gz{f}\trns \cdot (\bullet)^\flat
    && \text{and} &&
    \gz{\varphi}^{-1}_\star (\bullet)^\flat =  \gz{F}\trns \cdot (\bullet)^\flat \, , \\
    \gz{\varphi}_\star (\bullet)^\natural =  \gz{F} \cdot (\bullet)^\natural
    && \text{and} &&
    \gz{\varphi}^{-1}_\star (\bullet)^\natural =  \gz{f} \cdot (\bullet)^\natural  \, .
\end{align*}
Hence, for the covariant (kinematic) measures adopted here,
\begin{align*}
    \gz{e} =  \gz{\varphi}_\star (\gz{E}) && \text{and} &&  \gz{E} = \gz{\varphi}^{-1}_\star (\gz{e}) \, , \\
    \gz{i} =  \gz{\varphi}_\star (\gz{C}) && \text{and} &&  \gz{C} = \gz{\varphi}^{-1}_\star (\gz{i}) \, , \\
    \gz{c} =  \gz{\varphi}_\star (\gz{I}) && \text{and} &&  \gz{I} = \gz{\varphi}^{-1}_\star (\gz{c}) \, .
\end{align*}

\subsubsection{The micromorphic problem}

 The body $\mcl B$ is modelled as a micromorphic continuum to account for size-dependent effects.
As such, each material point $\mcl P \in \mcl B$ has additional micromorphic degrees of freedom associated with the continuum at the microscale that undergoes an affine deformation.
The micro-deformation $\mF(\X,t)$ represents an affine map of material points from their reference position $\overline{\X}$ to a current position $\overline{\x}$ within the microscale continuum; that is
\begin{align*}
    \overline{\x} = \mF \cdot \overline{\X} \, .
\end{align*}
The micro-deformation $\mF$ is kinematically independent of the macroscopic continuum and represents an additional state variable.
The gradient of the micro-deformation with respect to the macroscale material placement is a (mixed-variant) third-order tensor defined by
\begin{align*}
    \mG (\X) := \Grad  \mF (\X)
        = \dfrac{\partial \overline{F}_{iJ}}{\partial X_K} \gz{e}_i \otimes \gz{E}_J \otimes \gz{E}_K
        = \overline{G}_{iJK} \gz{e}_i \otimes \gz{E}_J \otimes \gz{E}_K \, .
\end{align*}

\subsubsection{The electric problem}

The scalar electric potential is denoted by $\potential$.
The spatial and referential electric fields, denoted by $\efield$ and $\Efield$ respectively, are thus given by
\begin{align*}
    \efield = - \grad \potential
    && \text{and} &&
    \Efield = - \Grad \potential \, ,
    \intertext{and are related to one another as follows}
    \efield = \gz{f}\trns \cdot \Efield = \gz{\varphi}_\star (\Efield)
    && \text{and} &&
    \Efield = \gz{F}\trns \cdot \efield = \gz{\varphi}^{-1}_\star (\efield) \, .
\end{align*}
The gradient operator with respect to the current configuration is defined by $\grad(\bullet) :=  \Grad(\bullet) \cdot \gz{f}$.

\subsection{Dirichlet principle (stationary energy principle)} \label{sec_dirichlet_principle}

The Dirichlet principle is employed to determine the kinetic quantities conjugate to the kinematic measures derived in \sect{sect_kinematics}.
The principle also provides the structure for the governing equations and the boundary conditions.

For the conservative system considered here, the total potential energy $E$ is given by
\begin{align}
E = \int_{\mcl B_0} U_0(\motion, \F, \mF, \mG, \potential, \Efield; \X) \, \d V
+
\int_{\Gamma_0} u_0(\motion, \mF, \potential; \X) \, \d A
\,,
\label{eq_total_potential_energy_1}
\end{align}
where $U_0$ and $u_0$ are the potential energy density functions per unit reference volume and area, respectively.
The potential energy density $U_0$ is additively decomposed as follows
\begin{align*}
    U_0 (\motion, \F, \mF, \mG, \potential, \Efield) = W_0(\F, \mF, \mG, \Efield) + V_0(\motion, \potential) \, ,
\end{align*}
where $W_0$ is the internal contribution and $V_0$ is the external contribution.
Note, we have assumed that there are no external contributions associated with the micromorphic problem.
Under the isothermal conditions assumed here, the internal contribution $W_0$ is further decomposed as
\begin{align}
W_0(\F, \mF, \mG, \Efield) =
\underbrace{
\psi_0^{\text{elast}}(\F, \mF, \mG, \Efield)
+
\psi_0^{\text{flexo}}(\{ \F,  \mF \} , \mG, \Efield)
}_{\psi_0(\F, \mF , \mG, \Efield)}
+
E_0(\F, \Efield) \, , \label{eq_W0}
\end{align}
where $\psi_0^{\text{elast}}$ is the electric free enthalpy density, $\psi_0^{\text{flexo}}$ is the internal energy associated with the flexoelectric effect, and $E_0$ is the electric energy density.
We note that the proposed additive decomposition is an assumption motivated by classical approaches in electroelasticity \citep[see e.g.][and references therein]{Vu2007}; other choices are possible.

\begin{rmk} \label{rmk_flex_energy_param}
Note, the internal energy associated with the flexoelectric effect describes the coupling between the gradient of the micro-deformation $\mG$ and the spatial electric field $\efield = \gz{f}\trns \cdot \Efield$.
Hence the dependence of the energy $\psi_0^{\text{flexo}}$ on both $\F$ and $\Efield$.
An alternative push-forward of $\Efield$ via $\overline{\gz{f}}:= \mF^{-1}$, is discussed further in \sect{sec_const_relations}.
The possible functional dependence of the internal energy associated with the flexoelectric effect on either $\F$ or $\mF$, or both, is denoted by curly braces in the parametrisation.
\qed
\end{rmk}

In summary, the total potential energy \eqref{eq_total_potential_energy_1} can be expressed as
\begin{equation}
    \begin{split}
    E   &= \int_{\mcl B_0} W_0(\F, \mF, \mG, \Efield; \X) \, \d V
        + \int_{\mcl B_0} V_0(\motion, \potential; \X) \, \d V
        + \int_{\Gamma_0} u_0(\motion, \mF, \potential; \X) \, \d A
    \, .
    \end{split}
    \label{eq_total_potential_energy_2}
\end{equation}
At equilibrium, the total potential energy of the system must be stationary with respect to arbitrary variations of the primary fields; that is
\begin{align*}
\delta  E(\motion, \F, \mF, \mG, \potential, \Efield) = 0 \, .
\end{align*}
Hence
\begin{equation}
    \begin{split}
    0 &= \int_{\mcl B_0}
    \left[
    \Ptot : \delta \F
    +\mP : \delta \mF
    +\mQ \tdot \delta \mG
    -\eD \cdot \delta \Efield
    -\gz b_0 \cdot \delta \gz \varphi
    +\rhob \delta \potential
    \, \d V
    \right] \\
&\quad
+ \int_{\Gamma_0}
\left[
-\t \cdot \delta \motion
-\mt : \delta \mF
+\rhos \delta \potential
\, \d A
\right]
\qquad \qquad
\forall \;
\delta \motion,\,
\delta \mF,\,
\delta \potential
\, ,
    \end{split} \label{eq_stationary_2}
\end{equation}
where the energetically-conjugate kinetic measures are defined in Table \ref{table_kinetics}.

\begin{table}[htb!]
    \centering
    \begin{tabular}{ l l l l}
     \toprule
    \textbf{Measure} &  \textbf{Domain}   & \textbf{Label} & \textbf{Order}   \\
     \hline
     $\Ptot := \D_{\F} U_0$ & $\mcl B_0$ & macroscopic Piola stress & 2  \\
     $\mP := \D_{\mF} U_0$  & $\mcl B_0$               & micromorphic Piola stress & 2  \\
    % $\Pmesh := \D_{\F} A^\text{mesh}_0$ &  $\mcl S_0$  & artificial Piola stress & 2   \\
    $\mQ := \D_{\mG} U_0$  & $\mcl B_0$ & micromorphic double stress & 3   \\
    $\eD := - \D_{\Efield} U_0$ & $\mcl B_0$  & dielectric displacement & 1  \\
    $\b_0 := - \D_{\motion} U_0$ & $\mcl B_0$   & body force & 1 \\
    $\rhob := \D_{\potential}U_0$ & $\mcl B_0$   & density of free charge per unit volume &
    0 \\
    $\t := -\D_{\motion} u_0$ & $\Gamma_0$ & macroscopic Piola traction & 1 \\
    $\mt := -\D_{\mF} u_0$ & $\Gamma_0$   & micromorphic Piola traction & 2 \\
    $\rhos := \D_{\potential} u_0$ & $\Gamma_0$    & density of free charge per unit area & 0 \\
    \bottomrule
    \end{tabular}
    \caption{
        Summary and definition of the kinetic measures introduced in \eqn{eq_stationary_2}.
        Order refers to the order of the tensorial quantity.
    }
    \label{table_kinetics}
    \end{table}

It is convenient to additively decompose the macroscopic Piola stress $\Ptot$ and the dielectric displacement $\eD$ as follows:
\begin{align*}
    \Ptot =:
    \underbrace{\left[
        \P + \Ppol  \right]}_{\D_{\F} \psi_0}
    + \underbrace{\left[ \Pmax \right]}_{\D_{\F} E_0}
    && \text{and} &&
    \eD =: \underbrace{\left[\eP  \right]}_{-\D_{\Efield} \psi_0}
    + \underbrace{\left[ \eD^\epsilon \right]}_{-\D_{\Efield} E_0} \, ,
\end{align*}
where $\psi_0 = \psi_0^{\text{elast}} + \psi_0^{\text{flexo}}$ was defined in \eqn{eq_W0}.
Here  $\P$ is the ordinary Piola stress,
$\Ppol$ is the polarization stress,
and $\Pmax$ is the Maxwell stress.
$\eD$ is the referential dielectric displacement,
$\eP$ is the referential  polarization,
and $\eD^\epsilon$ is the dielectric displacement.

The elastic contribution to the energy density associated with matter $\psi^\text{elast}_0$ (see \eqn{eq_W0}) contains contributions from the macroscopic problem, the micromorphic problem and an additional scale-bridging contribution \citep[see][for extensive details]{Hirschberger2008}; that is
\begin{align}
\psi_0^\text{elast}(\F, \mF, \mG, \Efield) =:
\psi_0^\text{mac} +
\psi_0^\text{mic} +
\psi_0^\text{scale} \, .
\label{eq_elastic_energy}
\end{align}

It is convenient to define the spatial polarization by $\mathbbm{p} := - D_\mathbbm{e} \psi_t$, where $\psi_t:= j \psi_0$ is the free enthalpy per unit volume of the current configuration, as the Piola transformation of the material polarization $\eP = -\D_{\Efield} E_0 $, that is
\begin{align*}
    \mathbbm{p} = j \gz{\varphi}_\star (\eP)
    && \text{and} &&
    \eP = J \gz{f} \cdot \mathbbm{p} = \mathbbm{p} \cdot \cof{\gz{F}} =   J \gz{\varphi}^{-1}_\star (\mathbbm{p}) \, ,
\end{align*}
where the cofactor of an invertible second-order tensor $(\bullet)$ is defined by $\cof (\bullet) := [\det (\bullet)] (\bullet)\invtrns$.

\begin{rmk}
     The electric energy density $E_0$ is parametrised here in terms of the electric field $\Efield$ and the deformation gradient $\F$.
     This is a common choice \citep[see][for further details]{Dorfmann2014}.
    \citet{Yvonnet2017} in their work on flexoelectricity uses a mixed-type formulation where the electric energy density $E_0$ is parametrised by the spatial polarization $\mathbbm{p}$.
    \qed
\end{rmk}

\subsection{Governing equations and boundary conditions}

The system of equations and boundary conditions governing the coupled problem of flexoelectricity and micromorphic elasticity are now derived.

The system of coupled governing equations (the Euler equations) is obtained by applying  the divergence theorem to \eqn{eq_stationary_2} and invoking the arbitrariness and independence of the variations $\delta \motion,\, \delta \mF$ and $\delta \potential$, to obtain
\begin{gather*}
    \left. \begin{split}
    \Div \left[ \P + \Ppol + \Pmax \right] + \gz b_0 = \gz 0  \\
    \Div \mQ - \mP = \gz 0 \\
    \Div \eD =  \rhob
    \end{split} \qquad \right\} \qquad \text{in } \mcl B_0 \, .
\end{gather*}

Dirichlet conditions on the displacement $\motion$, the micro-deformation $\mF$, and the electric potential $\potential$ are prescribed on the parts of the boundary $\Gamma^{\motion}_0 \subseteq \Gamma_0$, $\Gamma^{\mF}_0 \subseteq \Gamma_0$, and $\Gamma^{\potential}_0 \subseteq \Gamma_0$, respectively.
That is
\begin{align*}
    \motion = \motion^\text{pre}_{\Gamma} \text{ on } \Gamma^{\motion}_0 \, ,
    &&
    \mF =  \mF^\text{pre}_{\Gamma} \text{ on }  \Gamma^{\mF}_0 \, ,
    &&
    \potential_{\mcl B} =  \potential^\text{pre}_{\Gamma} \text{ on }  \Gamma^{\potential}_0 \, .
\end{align*}
The superscript $(\bullet)^\text{pre}$ denotes a prescribed function.
\begin{rmk}
    The physical meaning of a Dirichlet boundary condition on the micro-deformation $\mF$ is not clear.
    We retain this possibility for the sake of completeness.
    \qed
\end{rmk}

The various Neumann conditions on the respective subsets of $\Gamma_0$ follow as
\begin{align}
    \left[ \Pmax + \P + \Ppol \right]  \cdot \gz N  &= \t^\text{pre}
    && \text{on } \Gamma^{\P}_0
    \text{ where } \Gamma^{\motion}_0 \cup \Gamma^{\P}_0 = \Gamma_0 \text{ and } \Gamma^{\motion}_0 \cap \Gamma^{\P}_0 = \emptyset \, ,
    \label{neumann_motion} \\
    \mQ \cdot \gz N &= \mt^\text{pre}
    && \text{on } \Gamma^{\mQ}_0
    \text{ where }
    \Gamma^{\mF}_0 \cup \Gamma^{\mQ}_0 = \Gamma_0 \text{ and } \Gamma^{\mF}_0 \cap \Gamma^{\mQ}_0 = \emptyset\, ,
    \label{neumann_micro} \\
    -\left[ \eD^\epsilon + \eP\right]  \cdot \gz N &= \rhos{}^\text{pre}
    && \text{on } \Gamma^{\eD}_0
    \text{ where }
    \Gamma^{\potential}_0 \cup \Gamma^{\eD}_0 = \Gamma_0  \text{ and } \Gamma^{\potential}_0 \cap \Gamma^{\eD}_0 = \emptyset\, . \label{neumann_potential}
\end{align}

\subsection{Constitutive relations} \label{sec_const_relations}

Concrete examples for the various terms that comprise the total potential energy $E$ in \eqn{eq_total_potential_energy_2} are now given.
The resulting expressions for the kinetic measures defined in Table \ref{table_kinetics} are given in \ref{appendix_linearisation}.

\subsubsection{Elastic energy density}

Following \citep{Hirschberger2007, Pelteret2016}, the elastic energy density associated with matter
$\psi^\text{elast}_0 = \psi_0^\text{mac} +
\psi_0^\text{mic} +
\psi_0^\text{scale}
$
(see \eqn{eq_elastic_energy}) is assumed to be of the form
\begin{align}
    \psi_0^\text{mac}(\F, \Efield)
        &\equiv
        \frac{1}{2} \lambda \ln^2 J
        + \frac{1}{2} \mu \left[ \F : \F - n^\text{dim} - 2 \ln J \right]
        + \epsilon_0 \left[ \alpha  \gz I  + \beta  \gz C + \gamma \gz B \right ]
            : \Efield \otimes \Efield
        \, ,
        \label{psi_mac} \\
    \psi_0^\text{mic}(\mG)
        &\equiv
        \frac{1}{2} \mu \ell^2 \mG \tdot \mG \, ,
        \label{psi_mic} \\
    \psi_0^\text{scale}(\motion, \mF)
        &\equiv
        \frac{1}{2} p
        \left[ \mF - \F \right] :
        \left[ \mF - \F \right]
        \, .
        \label{psi_scale}
\end{align}
Here $\lambda$ and $\mu$ are the Lame constants, $\ell \geq 0$ is the length-scale parameter and $p \geq 0$ is a penalty-like parameter that couples the macro- and micro-deformation gradients.
The free space electric permittivity constant $\epsilon_0 = $\SI{8.854187817e-12}{\F \per \metre} and $\alpha$, $\beta$ and $\gamma$ are parameters.
\eqn{psi_mac} is an additive decomposition of a compressible neo-Hookean energy and a prototypical coupled electro-elastic model \citep[see][for further details]{Mehnert2018, Pelteret2016}.
Following \citet{Hirschberger2007}, the micromorphic and scale-bridging energies as chosen to be quadratic functions of the various strain measures.
We note that this is an assumption and not a requirement.

\begin{rmk}
    Alternative forms for the scale transition energy include
    \begin{align*}
        \psi_0^\text{scale}(\motion, \mF)
        &\equiv
        \frac{1}{2} p
        \left[ \invF \cdot \mF  - \gz{I} \right]^2
        \, , \\
        \psi_0^\text{scale}(\motion, \mF)
        &\equiv
        \frac{1}{2} p
        \left[ \mF\trns \cdot \mF  - \gz{C} \right]^2 \, .
    \end{align*}
    An alternative description of the micromorphic energy density in terms of Eringen's Lagrangian micro-deformation gradient $\overline{\gz G}{}^\text{E} := \mF \trns \cdot \mG$ is
    \begin{align*}
        \psi_0^\text{mic}(\mF,\mG)
        &\equiv \frac{1}{2} \mu \ell^2 \overline{\gz G}{}^\text{E}  \tdot \overline{\gz G}{}^\text{E}  \, .
    \end{align*}
\end{rmk}

\subsubsection{Electric energy density}

The electric energy density is given by \citep[see][]{Vu2006}
\begin{align}
    E_0(\F, \Efield) = -\dfrac{1}{2} \epsilon_0 J  \, \gz{B} : \Efield \otimes \Efield
 \, .
\label{eq_hyperelastic_energy_electric}
\end{align}

\subsubsection{Flexoelectric energy density}

As discussed in \sect{sec_dirichlet_principle}, the flexoelectric contribution couples the third-order, mixed-variant micro-gradient $\mG$ and the electric field.
We propose here that the flexoelectric contribution takes the form
\begin{align}
    \psi_0^\text{flexo}(\mF, \mG, \Efield)
    &=
    \upsilon \left[ \invmF \trns  \cdot \Efield  \right] \cdot \mG : \gz{I}
    \, ,
    \label{psi_flexo_A}
\end{align}
where $\upsilon := \epsilon_0 \ell \overline{\upsilon}$ is a positive parameter.
The inclusion of the length scale $\ell$ increases the relative flexoelectric contribution for diminishing sample size (cf.\ \eqn{psi_mic}).
The free space electric permittivity constant ensures that the contribution of the flexoelectric energy is of a similar order and structure to the electrical contributions in $\psi_0^\text{mac}$ and $E_0$, see \eqn{psi_mac} and \eqn{eq_hyperelastic_energy_electric}, respectively.
The pull-back of $\Efield$ via the micro-deformation is proposed to preclude direct coupling of the macroscale Piola stress $\Ptot = \D_{\F} U_0$ and the microscale problem other than through the scale-bridging energy in \eqn{psi_scale} (see Remark \ref{rmk_flex_energy_param}).

\begin{rmk}
An alternative form of \eqn{psi_flexo_A} would be
\begin{align*}
    \psi_0^\text{flexo}(\F, \mG, \Efield)
    =
    \upsilon \,  \efield \cdot \mG : \gz{I}
    =
    \upsilon \left[ \gz{\varphi}_\star (\Efield) \right] \cdot \mG :  \gz{\varphi}^{-1}_\star (\gz{b})
    =
    \upsilon \left[ \invF\trns \cdot \Efield \right] \cdot \mG :  \gz{I} \, .
    % \label{psi_flexo_B}
\end{align*}
We note that $\mG$ is a mixed-variant tensor that is contracted in $\psi_0^\text{flexo}$ from the left by the covariant spatial electric field $\mathbbm{e} = \gz{\varphi}_\star (\Efield)$ and from the right by the contravariant material identity tensor $\gz{I}$.

In the same spirit, a further logical proposal for the flexoelectric energy would be
\begin{align*}
    \psi_0^\text{flexo}(\F, \mG, \Efield)
    =
    \upsilon \,  \efield \cdot \mG : \gz{B}
    =
    \upsilon \left[ \gz{\varphi}_\star (\Efield) \right] \cdot \mG :  \gz{\varphi}^{-1}_\star (\gz{i}) \, .
    % \label{psi_flexo_C}
\end{align*}
\qed
\end{rmk}

\begin{rmk}
    The model of flexoelectricity proposed by \citet{Yvonnet2017} requires a $C^1$-continuous finite element approximation.
    This is restrictive.
    The micromorphic approach proposed here requires standard $C^0$ continuity.
    To compare formulations, define the gradient of the deformation gradient by $\gz G = \Grad \F = \Grad [ \Grad \motion ]$.
    We note that as the penalty-like parameter $p \to \infty$ in \eqn{psi_scale}, $\mG \to \gz G$.
    \citeauthor{Yvonnet2017} define an internal energy to describe the gradient and the flexoelectric effect that takes the form
    \begin{align}
        \psi_0^\text{flexo}(\gz G, \mathbbm{p}) = \frac{1}{2}[\ell^\text{YL}]^2
        \left[
            \gz G : \gz I
        \right] \cdot \left[
            \gz G : \gz I
        \right]
        +
        \upsilon^\text{YL} \,  \mathbbm{p} \cdot  \left[ \gz G : \gz I \right]
     \, ,
     \label{psi_flex_JY}
    \end{align}
    where $\ell^\text{YL} \geq 0$ is a length scale and $\upsilon^\text{YL} \geq 0$ is a constitutive parameter.
    Note that the units of $\upsilon$ and $\upsilon^\text{JY}$ are clearly different.
    The first term in \eqn{psi_flex_JY} is similar to the micromorphic energy in \eqn{psi_mic} but with a different choice of inner product.
    The second term is similar but reflects the choice of  \citeauthor{Yvonnet2017} to select the polarization $\mathbbm{p}$ as a primary field.
    \qed
\end{rmk}

\begin{figure}[htb!]
    \centering
    \includegraphics[width=0.85\textwidth]{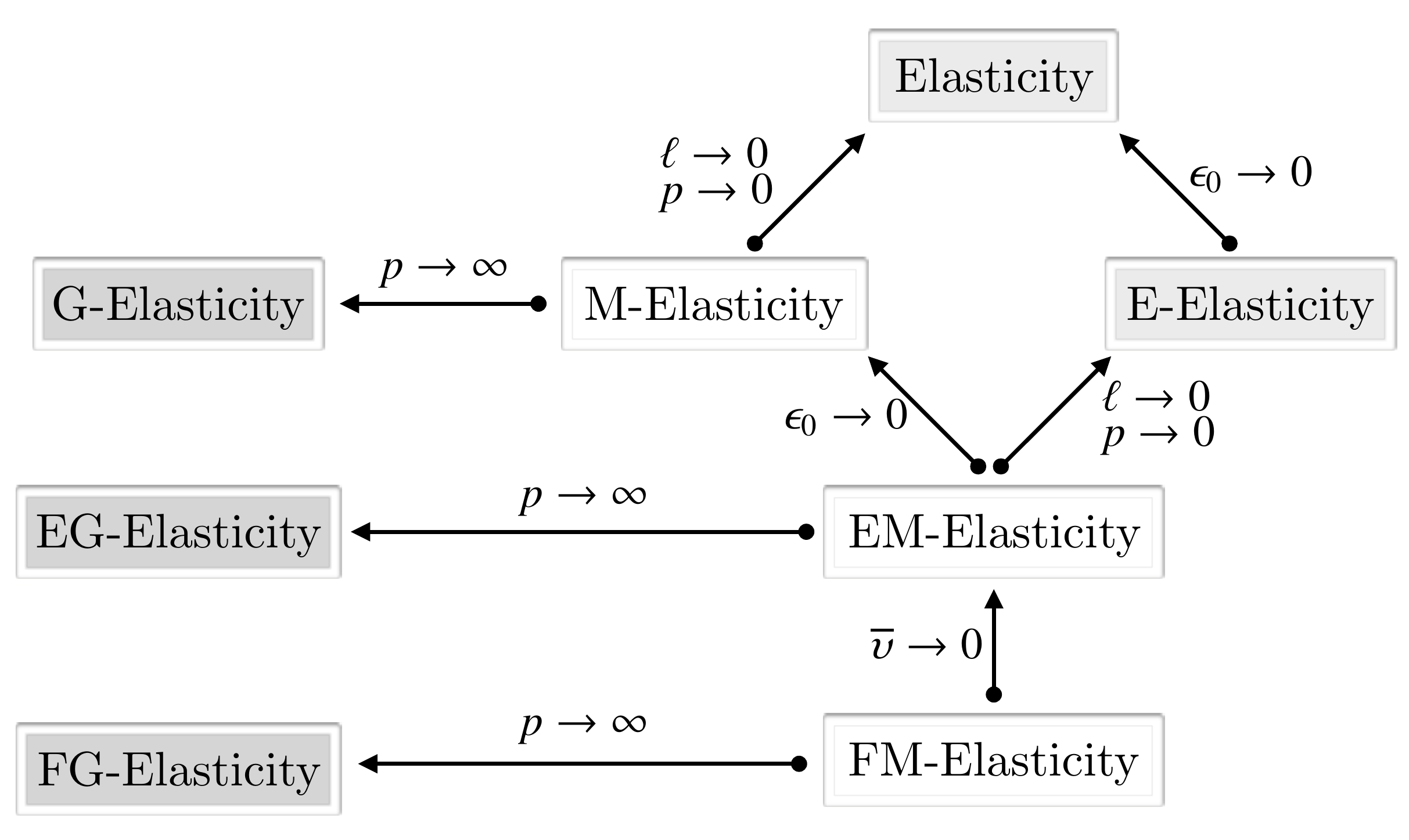}
    \caption{The relation between the various  models of coupled electrical and mechanical elasticity within the micromorphic setting.}
    \label{fig_family_tree}
  \end{figure}

The micromorphic model of flexoelectricity allows a range of different problems to be addressed by modifying the parameters in the constitutive relations,  as depicted in \fig{fig_family_tree}.
The schematic provides a convenient classification structure.
FM-Elasticity denotes the problem of coupled flexoelectricity and micromorphic elasticity.
As the penalty penalty-like parameter $p \to \infty$ we recover coupled flexoelectricity and gradient elasticity, denoted FG-Elasticity.
By setting the flexoelectric parameter $\overline{\upsilon} \to 0$ we obtain the problem of coupled electro-micromorphic elasticity, denoted EM-Elasticity.
Note that by this definition, flexoelectric  effects are absent in EM-Elasticity.
The problem of coupled gradient electro-elasticity is obtained from  EM-Elasticity as $p \to \infty$.
In addition micromorphic elasticity (M-Elasticity) and electro-elasticity (E-Elasticity) are obtained from EM-Elasticity as $\epsilon_0 \to 0$, and as $\ell \to 0$ and $p \to 0$, respectively.
In the same spirit, we recover gradient elasticity from M-Elasticity as $p \to \infty$.
Finally, we obtain the standard problem of nonlinear elasticity from M-Elasticity as $\ell \to 0$ and $p \to 0$, and from E-elasticity as $\epsilon_0 \to 0$.

Henceforth for the choice of $\ell \equiv 0$ it is implied that $p \equiv 0$.
This ensures that the macroscopic and micromorphic response are uncoupled.

\section{The Finite Element approximation} \label{sec_fe_approximation}

The triangulation of the reference configuration $\mcl B_0$ into non-overlapping elements is denoted by $\mathcal{T}^h_{\mcl B}$.
The primary fields (the macroscopic motion $\motion$, the micro-deformation $\overline{\gz F}$, and the electric potential $\potential$) are approximated using finite element spaces of continuous piecewise polynomials of fixed, but potentially different,  degree.
The macroscopic motion $\gz \varphi \in H^1(\mcl B_0)$, the micromorphic deformation $\overline{\gz F} \in H^1(\mcl B_0)$, and the scalar electric potential $\varphi \in H^1(\mcl B_0)$ are respectively given in a vector space spanned by the standard (i.e.\ $C^0$-continuous) vector-, tensor-, and scalar-valued finite element basis functions (polynomials with local support), respectively denoted by $\gz N^I_{\gz \varphi}$, $\gz N^I_{\mF}$ and $N^I_{\varphi}$.
That is, the primary fields and their associated variations ($\delta \gz \varphi \in H^1_0(\mcl B_0)$, $\delta  \mF \in H^1_0(\mcl B_0)$ and $\delta \varphi \in H^1_0(\mcl B_0)$) are approximated by
\begin{align}
       \gz \varphi^h =:
       \sum_{I \in \mcl I_{\gz \varphi}}  {\gz \upvarphi}_I \gz N^I_{\gz \varphi} (\gz X)
        &&  \text{and} &&
       \delta \gz \varphi^h =:  \sum_{I \in \mcl I_{\gz \varphi}}  {\gz \upvarphi}_I \gz N^I_{\gz \varphi} (\gz X)
       \label{eq_motion_h} \,, \\
       \mF^h =:  \sum_{I \in \mcl I_{\mF}}  \overline{\mathbf{\mathsf{F}}}_I \gz N^I_{\mF} (\gz X)
        && \text{and}  &&
       \delta \mF^h =:  \sum_{I \in \mcl I_{\mF}}  \delta \overline{\mathbf{\mathsf{F}}}_I \gz N^I_{\mF} (\gz X)
       \label{eq_mF_h} \,, \\
       \varphi^h =:  \sum_{I \in \mcl I_{\varphi}}  {\upvarphi}_I  N^I_{\varphi} (\gz X)
        && \text{and} &&
       \delta  \varphi^h =:  \sum_{I \in \mcl I_{\varphi}}  \delta {\upvarphi}_I N^I_{\varphi} (\gz X)
       \label{eq_epot_h}
       \, ,
\end{align}
where superscript $h$ indicates that the representation is related to the finite element mesh with size function $h(\gz X)$.
Upright Greek letters are used to denote a global vector containing the degrees of freedom associated with one of the three primary field.
The sets $\mcl I_{\motion}$ and $\mcl I_{\mF}$ and $\mcl I_{\varphi}$ contain the degrees of freedom for the
macroscopic, micromorphic and electric fields, respectively.
The discrete representation of the gradients  and variations of the primary fields follow directly as
\begin{align}
    \F^h =:  \sum_{I \in \mcl I_{\gz \varphi}}  {\gz \upvarphi}_I \Grad \gz N^I_{\gz \varphi} (\gz X)
    &&  \text{and} &&
    \delta \F^h =:  \sum_{I \in \mcl I_{\gz \varphi}} \delta {\gz \upvarphi}_I \Grad \gz N^I_{\gz \varphi} (\gz X)
     \,, \\
    \mG^h =:  \sum_{I \in \mcl I_{\mF}}  \overline{\mathbf{\mathsf{F}}}_I \Grad \gz N^I_{\mF} (\gz X)
    && \text{and}  &&
    \delta \mG^h =:  \sum_{I \in \mcl I_{\mF}}  \delta \overline{\mathbf{\mathsf{F}}}_I \Grad \gz N^I_{\mF} (\gz X)
     \,, \\
    \Efield^h =:  -\sum_{I \in \mcl I_{\varphi}}  {\upvarphi}_I \Grad  N^I_{\varphi} (\gz X)
    && \text{and} &&
    \delta  \Efield^h =:  -\sum_{I \in \mcl I_{\varphi}}  \delta {\upvarphi}_I \Grad  N^I_{\varphi} (\gz X)
    \, . \label{eq_Efield_h}
\end{align}

Substituting the discrete representations \eqref{eq_motion_h}--\eqref{eq_Efield_h} into the stationary condition \eqref{eq_stationary_2}, yields the following three sets of coupled non-linear residual equations to be satisfied:
\begin{align}
    %R1
    \mathsf{R}^{I}_{\motion} &:=
    \int_{\mcl B_0}
    \left[
    {\Ptot} : \Grad \gz N^I_{\motion} -
    \gz b_0 \cdot \gz N^I_{\motion}
    \right] \, \d V
    - \int_{\Gamma_0} \t \cdot \gz N^I_{\motion} \, \d A
    \doteq 0
    &&
    \forall I \in \mcl I_{\motion}
    \label{R_motion}
    \\
    %R2
    \mathsf{R}^{I}_{\mF} &:=
    \int_{\mcl B_0}
    \left[
    \mP : \gz N^I_{\mF} +
    \mQ \tdot\Grad \gz N^I_{\mF}
    \right] \, \d V
    - \int_{\Gamma^{\mcl B}_0}  \mt : \gz N^I_{\mF} \, \d A
    \doteq 0
    &&
    \forall I \in \mcl I_{\mF}
    \label{R_mF}
    \\
    \mathsf{R}^{I}_{\potential} &:=
    \int_{\mcl B_0}
    \left[
     \eD \cdot \Grad N^I_{\potential}
     + \rhob N^I_{\potential}
    \right] \, \d V
    + \int_{\Gamma_0} \rhos N^I_{\potential} \, \d A
    \doteq 0
    &&
    \forall I \in \mcl I_{\potential} \, .
    \label{R_potential}
\end{align}

The three global residual vectors, obtained by assembling the individual  contributions from the residual expressions associated with the respective degrees of freedom \eqref{R_motion}--\eqref{R_potential}, are denoted by
\begin{align*}
    \begin{bmatrix}
        \gz{\mathsf{R}}_{\motion} &
        \gz{\mathsf{R}}_{\mF}     &
        \gz{\mathsf{R}}_{\varphi}
    \end{bmatrix}\trns
    =: \gz{\mathsf{R}} \, .
\intertext{and the global vectors of degrees of freedom by}
    \begin{bmatrix}
        \gz{\mathsf{d}}_{\motion} &
        \gz{\mathsf{d}}_{\mF}     &
        \gz{\mathsf{d}}_{\varphi}
    \end{bmatrix}\trns
    =: \gz{\mathsf{d}} \, .
\end{align*}
Note that
$\dim \gz{\mathsf{R}}_{\motion} = \dim \gz{\mathsf{d}}_{\motion} =  \vert \mcl I_{\gz \varphi} \vert $,
$\dim \gz{\mathsf{R}}_{\mF} = \dim \gz{\mathsf{d}}_{\mF} =  \vert \mcl I_{\gz \mF} \vert $, and
$\dim \gz{\mathsf{R}}_{\potential} = \dim \gz{\mathsf{d}}_{\potential} =  \vert \mcl I_{\gz \potential} \vert $.

The coupled nonlinear residual equations are solved approximately using a Newton--Raphson strategy whereby within each iteration $(i)$ of the current load (time) step the linearised problem is given by
\begin{gather*}
    \gz{\mathsf{R}}^{(i+1)}
    = \gz{\mathsf{R}}^{(i)} + \left[ \D_{\gz{\mathsf{d}}} \gz{\mathsf{R}}^{(i)} \right] \Delta \gz{\mathsf{d}}^{(i)} \doteq \gz 0 \\
    \implies
    \gz{\mathsf{K}}^{(i)} \Delta \gz{\mathsf{d}}^{(i)}
    =
    - \gz{\mathsf{R}}^{(i)} \, ,
\end{gather*}
and $\Delta \gz{\mathsf{d}}^{(i)} := \gz{\mathsf{d}}^{(i+1)} - \gz{\mathsf{d}}^{(i)}$.
When expressed in the form of a block system, the discrete problem at each iteration takes the form
\begin{align}
    \begin{bmatrix}
       \gz{\mathsf{K}}_{\motion\motion} &
       \gz{\mathsf{K}}_{\motion\mF} &
       \gz{\mathsf{K}}_{\motion\varphi}
       \\[6pt]
       \gz{\mathsf{K}}_{\mF\motion} &
       \gz{\mathsf{K}}_{\mF\mF} &
       \gz{\mathsf{K}}_{\mF\potential} &
       \\[6pt]
       \gz{\mathsf{K}}_{\potential\motion} &
       \gz{\mathsf{K}}_{\potential\mF} &
       \gz{\mathsf{K}}_{\potential\potential}
    \end{bmatrix}^{(i)}
    \begin{bmatrix}
        \Delta \gz{\mathsf{d}}_{\motion} \\[6pt]
        \Delta \gz{\mathsf{d}}_{\mF}     \\[6pt]
        \Delta \gz{\mathsf{d}}_{\potential}
    \end{bmatrix}^{(i)}
    =
    -
    \begin{bmatrix}
        \gz{\mathsf{R}}_{\motion} \\[6pt]
        \gz{\mathsf{R}}_{\mF}     \\[6pt]
        \gz{\mathsf{R}}_{\potential}
    \end{bmatrix}^{(i)} \, .
    \label{KD_R}
\end{align}
The load step is deemed converged when the (normalised) magnitude of the incremental changes
$\Delta \gz{\mathsf{d}}_{\motion}$,
$\Delta \gz{\mathsf{d}}_{\mF}$, and
$\Delta \gz{\mathsf{d}}_{\potential}$,
together with the (normalised) magnitude of the residual vectors
$\gz{\mathsf{R}}_{\motion}$,
$\gz{\mathsf{R}}_{\mF}$, and
$\gz{\mathsf{R}}_{\potential}$,
are below a defined tolerance $\epsilon \ll  1$.

The explicit reference to the current iteration counter is dropped henceforth.
The matrix problem \eqref{KD_R} is solved monolithically.
The various contributions to the tangent matrix $\gz{\mathsf{K}}$ associated with the degrees of freedom $I \in \{\mcl I_{\motion}, \mcl I_{\mF}, \mcl I_{\varphi} \}$ and $J \in \{\mcl I_{\motion}, \mcl I_{\mF}, \mcl I_{\varphi} \}$ are given by
\begin{align*}
    % 11
    {\left[ {\mathsf{K}_{\motion\motion}}\right]}_{IJ}
    &=
    \partial_{\smotion^J} {\mathsf{R}}^{I}_{\motion}
    =
    \int_{\mcl B_0} \partial_{\smotion^J} \left[ \Ptot : \Grad  \gz N^I_{\motion} \right] \, \d V
    =
    \int_{\mcl B_0} \left[ \D_{\F}{\Ptot} : \Grad  \gz N^J_{\motion}  \right]: \Grad  \gz N^I_{\motion} \, \d V
    \\
    % 12
    {\left[ {\mathsf{K}_{\motion\mF}}\right]}_{IJ}
    &=
    \partial_{\smF^J} {\mathsf{R}}^{I}_{\motion}
    =
    \int_{\mcl B_0} \partial_{\smF^J} \left[ \Ptot : \Grad  \gz N^I_{\motion} \right] \, \d V
    =
    \int_{\mcl B_0} \left[ \D_{\mF}{\Ptot} : \gz N^J_{\mF}  \right]: \Grad  \gz N^I_{\motion} \, \d V
    \\
    % 13
    {\left[ {\mathsf{K}_{\motion\varphi}}\right]}_{IJ}
    &=
    \partial_{\spotential^J} {\mathsf{R}}^{I}_{\motion}
    =
    \int_{\mcl B_0} \partial_{\spotential^J} \left[ \Ptot : \Grad  \gz N^I_{\motion} \right] \, \d V
    =
    - \int_{\mcl B_0} \left[ \D_{\Efield}{\Ptot} \cdot \Grad N^J_{\potential}  \right]: \Grad  \gz N^I_{\motion} \, \d V
    \\
    %-------------------------------------------------
    \cline{1-2}
    % 21
    {\left[ {\mathsf{K}_{\mF\motion}}\right]}_{IJ}
    &=
    \partial_{\smotion^J}  {\mathsf{R}}^{I}_{\smF}
    =
    \int_{\mcl B_0} \partial_{\smotion^J}  \left[
        \mP : \gz N^I_{\mF} +
        \mQ \tdot\Grad \gz N^I_{\mF}
        \right]
        \, \d V
    \\
    &=
    \int_{\mcl B_0} \left[ \D_{\F}{\mP} : \Grad \gz N^J_{\motion}  \right]: \gz N^I_{\mF} \, \d V
    +
    \int_{\mcl B_0} \left[ \D_{\F}{\mQ} \tdot \Grad \gz N^J_{\motion}  \right] \tdot \Grad \gz N^I_{\mF} \, \d V
    \\
    % 22
    {\left[ {\mathsf{K}_{\mF\mF}}\right]}_{IJ}
    &=
    \partial_{\smF^J}  {\mathsf{R}}^{I}_{\smF}
    =
    \int_{\mcl B_0} \partial_{\smF^J}  \left[
        \mP : \gz N^I_{\mF} +
        \mQ \tdot\Grad \gz N^I_{\mF}
        \right]
        \, \d V
    \\
    &=
    \int_{\mcl B_0} \left[ \D_{\mF}{\mP} : \gz N^J_{\mF}  \right]: \gz N^I_{\mF} \, \d V
    +
    \int_{\mcl B_0} \left[ \D_{\mG}{\mQ} \tdot \Grad \gz N^J_{\mF}  \right] \tdot  \Grad \gz N^I_{\mF} \, \d V
    \\
        % 23
        {\left[ {\mathsf{K}_{\mF\potential}}\right]}_{IJ}
        &=
        \partial_{\spotential^J}  {\mathsf{R}}^{I}_{\smF}
        =
        \int_{\mcl B_0} \partial_{\spotential^J}  \left[
            \mP : \gz N^I_{\mF} +
            \mQ \tdot\Grad \gz N^I_{\mF}
            \right]
            \, \d V
        \\
        &=
        - \int_{\mcl B_0} \left[ \D_{\Efield}{\mP} \cdot \Grad N^J_{\potential}  \right]: \gz N^I_{\mF} \, \d V
        -
        \int_{\mcl B_0} \left[ \D_{\mG}{\mQ} \cdot \Grad N^J_{\potential}  \right] \tdot  \Grad \gz N^I_{\mF} \, \d V
        \\
        %-------------------------------------------------
        \cline{1-2}
    %31
    {\left[ {\mathsf{K}_{\potential\motion}}\right]}_{IJ}
    &=
    \partial_{\smotion^J}  {\mathsf{R}}^{I}_{\spotential}
    =
    \int_{\mcl B_0} \partial_{\smotion^J}  \left[
        \eD \cdot \Grad N^I_{\potential}
        \right]
        \, \d V
    =
    \int_{\mcl B_0} \left[ \D_{\F}{\eD} : \Grad \gz N^J_{\motion}  \right] \cdot \Grad N^I_{\potential} \, \d V
    \\
    %32
    {\left[ {\mathsf{K}_{\potential\mF}}\right]}_{IJ}
    &=
    \partial_{\smF^J}  {\mathsf{R}}^{I}_{\spotential}
    =
    \int_{\mcl B_0} \partial_{\smF^J}  \left[
        \eD \cdot \Grad N^I_{\potential}
        \right]
        \, \d V
    =
    \int_{\mcl B_0} \left[ \D_{\mF}{\eD} : \gz N^J_{\mF}  \right] \cdot \Grad N^I_{\potential} \, \d V
    \\
    %33
    {\left[ {\mathsf{K}_{\potential\potential}}\right]}_{IJ}
        &=
        \partial_{\spotential^J}  {\mathsf{R}}^{I}_{\spotential}
        =
        \int_{\mcl B_0} \partial_{\spotential^J}  \left[
            \eD \cdot \Grad N^I_{\potential}
            \right]
            \, \d V
        =
        - \int_{\mcl B_0} \left[ \D_{\Efield}{\eD} \cdot \Grad  N^J_{\potential}  \right] \cdot \Grad N^I_{\potential} \, \d V
        \, .
\end{align*}

\section{Numerical examples} \label{sect_numerical_examples}

The finite element problem detailed in the previous section is implemented within the open-source library deal.II \citep{dealII91,Bangerth2007} in conjunction with the linear algebra package Trilinos \citep{Trilinos2005}.
The automatic differentiation package ADOL-C \citep{adolc2012} is used to evaluate the derivatives that appear in the expressions for the residual and tangent given in \sect{sec_fe_approximation}.

Tri-quadratic piecewise polynomials are used to approximate the (vectorial) macroscopic motion map $\gz \varphi^h$ and the (tensorial) micromorphic deformation gradient $\overline {\gz F}^h$.
 Tri-linear approximations are used for the (scalar) electric potential $\potential^h$.
This is a non-standard choice for the problem of E-Elasticity where equal-order approximations are typically used for $\gz \varphi^h$ and $\potential^h$ \citep{Pelteret2016}.
We choose to break with convention to ensure that the flexoelectric contribution to the energy in \eqn{psi_flexo_A} contains electrical and micromorphic terms of equal polynomial order.
The challenge of determining the optimal functional setting for the problem of FM-Elasticity is discussed further in \sect{sec_discuss_conclusion}.

The default constitutive parameters used for the numerical examples, unless stated otherwise, are listed in Table~\ref{table_constitutive_parameters}.
To improve the scaling of the linear system \eqref{KD_R}, we adopt units of \si{mm}, \si{N} and \si{\kilo \V}.
Initial conditions on the micro-deformation of $\mF \equiv \overline{\gz{I}}$ are set, where $\overline{\gz{I}}$ is the second-order micromorphic identity tensor.
Homogeneous Neumann boundary conditions for the micromorphic traction, defined in \eqn{neumann_micro}, are assumed for all example problems.
Body forces and free charge are ignored.

    \begin{table}[htb!]
        \centering
        \begin{tabular}{ l l l l l}
         \toprule
        \textbf{Parameter} &  \textbf{Symbol}   & \textbf{Value} \\
         \hline
         Poisson's ratio    & $\nu$     & \num{0.273}               \\
         Shear modulus      & $\mu$     & \num{0.05}       \\
                            & $\alpha$  & \num{0.2}                 \\
                            & $\beta$   & \num{2}                   \\
                            & $\gamma$   &  \num{-2}                   \\
        Penalty-like parameter & $p$    & $\num{5000}\mu$           \\
        \bottomrule
        \end{tabular}
        \caption{Default constitutive parameters.
        The electrical and mechanical parameters are taken from \citep{Vu2010, Vu2012, Pelteret2016}. }
        \label{table_constitutive_parameters}
        \end{table}

We consider two three-dimensional example problems to elucidate the theory developed in the previous sections.
The first is the problem of a strip with a hole, the second the bending of a cantilever beam.

\subsection{Strip with hole: M-Elasticity}\label{sec_strip_hole_m_elasticity}

The objective of this example is to demonstrate the key features of  M-Elasticity.
Specifically, we examine the role that the ratio of the length scale $\ell$ to a characteristic dimension of the macroscopic problem $L$ plays in the overall response of the structure.
Recall that in the proposed formulation for FM-Elasticity, the micromorphic model captures scale-dependent effects and allows the gradient of the deformation gradient $\gz{G} = \Grad \F$ to be approximated via its micromorphic counterpart $\mG$ within a conventional $C^0$-continuous finite element setting.

\begin{figure}[htb!]
    \centering
    \includegraphics[width=0.65\textwidth]{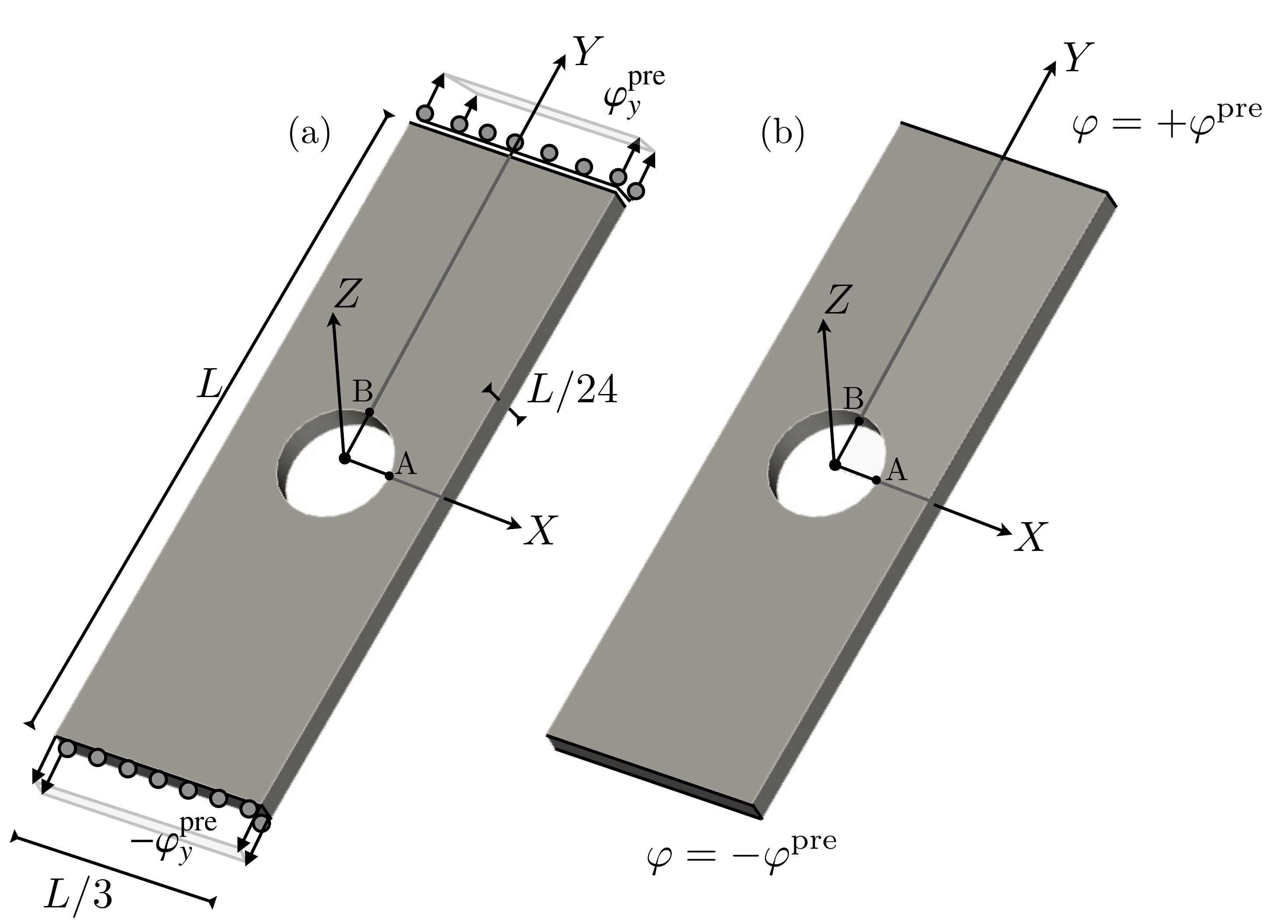}
    \caption{The problem of a strip with a hole.
    Geometry and boundary conditions for (a) M-Elasticity and (b) E- and FM-Elasticity.}
    \label{fig_strip_with_hole_geom_bcs}
  \end{figure}

\begin{figure}[htb!]
    \centering
    \includegraphics[width=\textwidth]{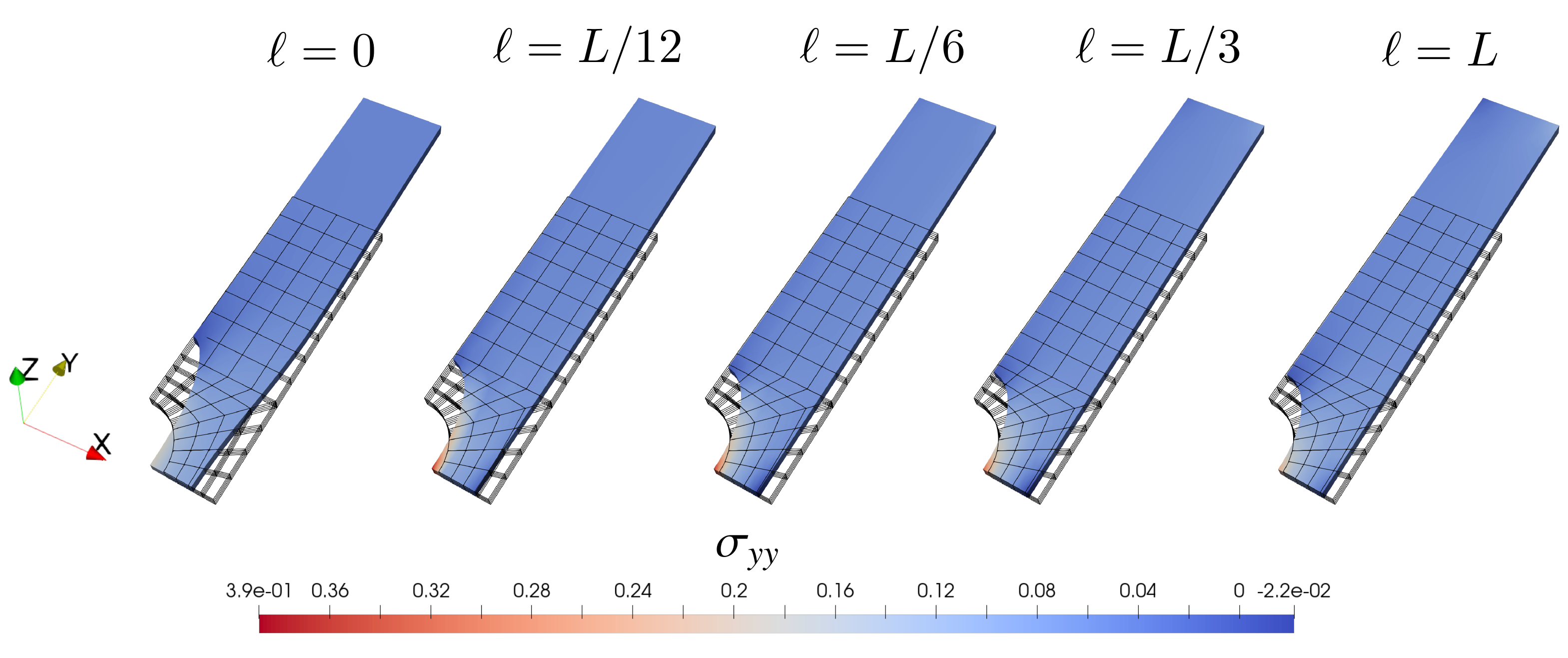}
    \caption{The finite element mesh of the undeformed strip with a hole geometry, and the final deformed configuration for various different length scales $\ell$.
    The distribution of the $yy$-component of the Cauchy stress $\gz{\sigma} := j \P  \cdot \F \trns$ is shown.
    }
    \label{fig_strip_hole_m_elasticity}
  \end{figure}

Consider the problem of a three-dimensional strip with dimensions $L \times L/3 \times L/24$, where $L=120$, loaded in tension as depicted in \fig{fig_strip_with_hole_geom_bcs}(a).
A two-dimensional version of the problem was proposed by \citet{Hirschberger2007} for the problem of M-Elasticity.
An equal and opposite motion is prescribed on the upper and lower face in 5 equal steps.
The final length of the deformed specimen is $3L /2$.
All other boundary conditions are of type homogeneous Neumann.
The finite element mesh of the undeformed strip is shown in \fig{fig_strip_hole_m_elasticity}.
The symmetry of the problem is exploited to reduce the computational cost by simulating only one quarter of the domain.
Following \citep{Hirschberger2007}, the penalty-like parameter is set to $p \equiv 50 \mu$ for this example.

\begin{figure}[htb!]
    \centering
    \includegraphics[width=\textwidth]{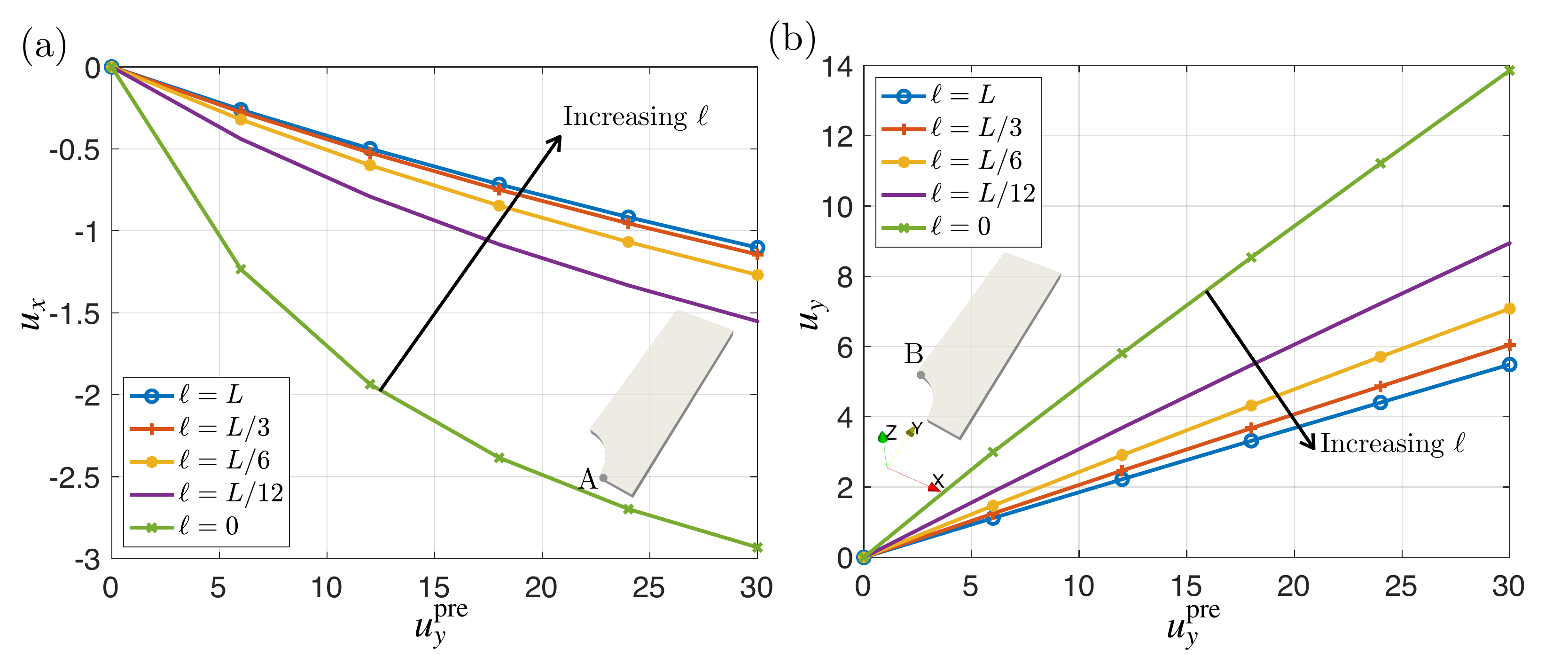}
    \caption{The relationship between the applied displacement $u_y^\text{pre}$ and the (a) horizontal displacement of point A and (b) the vertical displacement of point B for the strip with a hole problem and the model of M-Elasticity.}
    \label{fig_strip_hole_m_elasticity_plots}
  \end{figure}

The final deformed shapes for five different choices of $\ell \in \{0;L/12;L/6;L/3;L \}$ are shown in \fig{fig_strip_hole_m_elasticity_plots}.
The choice $\ell \equiv 0$ corresponds to (nonlinear) Elasticity, see \fig{fig_family_tree}.
The response away from the hole is similar for all choices of $\ell$ as the deformation is essentially homogeneous in this region.
The micromorphic effect is significant in the vicinity of the hole where the deformation is inhomogeneous.
The horizontal and the vertical displacement of the points A and B, respectively, (see \fig{fig_strip_with_hole_geom_bcs}(a)) are plotted against the prescribed displacement of the upper face $\varphi_y^\text{pre}$ in \fig{fig_strip_hole_m_elasticity_plots}.
Increasing the length scale, or equivalently reducing the specimen size, leads to a stiffer response.

The increase of strength with decreasing specimen size is the key feature of M-Elasticity.
This behaviour will be inherited by the problems of EM- and FM-Elasticity, as discussed next.

  \subsection{Strip with hole: E- and FM-Elasticity}

\begin{figure}[htb!]
    \centering
    \includegraphics[width=0.7\textwidth]{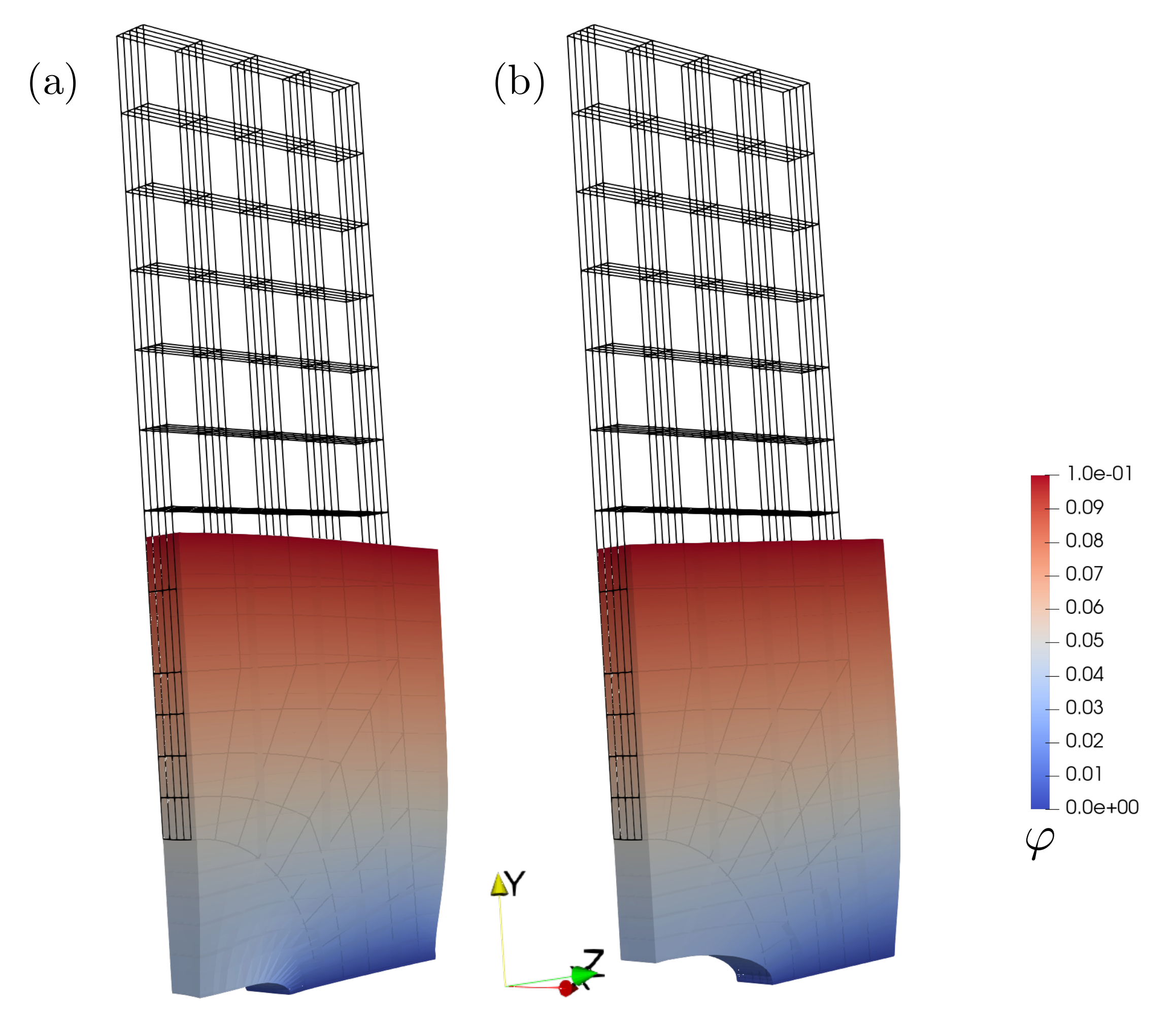}
    \caption{The finite element mesh of the undeformed configuration and the deformed strip for the problems of (a) E-Elasticity and (b) FM-Elasticity.
    The plot is coloured by the potential field.}
    \label{fig_strip_hole_e_fm_elasticity}
  \end{figure}

The problem of a strip with a hole has been used to demonstrate key features of models of E-Elasticity at finite deformations \citep[see e.g.][]{Vu2007}.
The objective here is to use this benchmark problem to illustrate differences between E- and FM-Elasticity.
The problem of EM-Elasticity is not considered in this example problem for the sake of brevity.
The boundary conditions are shown in \fig{fig_strip_with_hole_geom_bcs}(b).
A potential difference of $\num{0.2} \equiv 2 \potential^\text{pre}$ is applied between the upper and lower faces in five equal steps.
The finite element mesh of the undeformed configuration is shown in \fig{fig_strip_hole_e_fm_elasticity}.
As in \sect{sec_strip_hole_m_elasticity}, the symmetry of the problem is exploited.
The length scale is fixed as $\ell \equiv L/3$ and the penalty-like parameter is set to $p \equiv 5000 \mu$.

\begin{figure}[htb!]
    \centering
    \includegraphics[width=\textwidth]{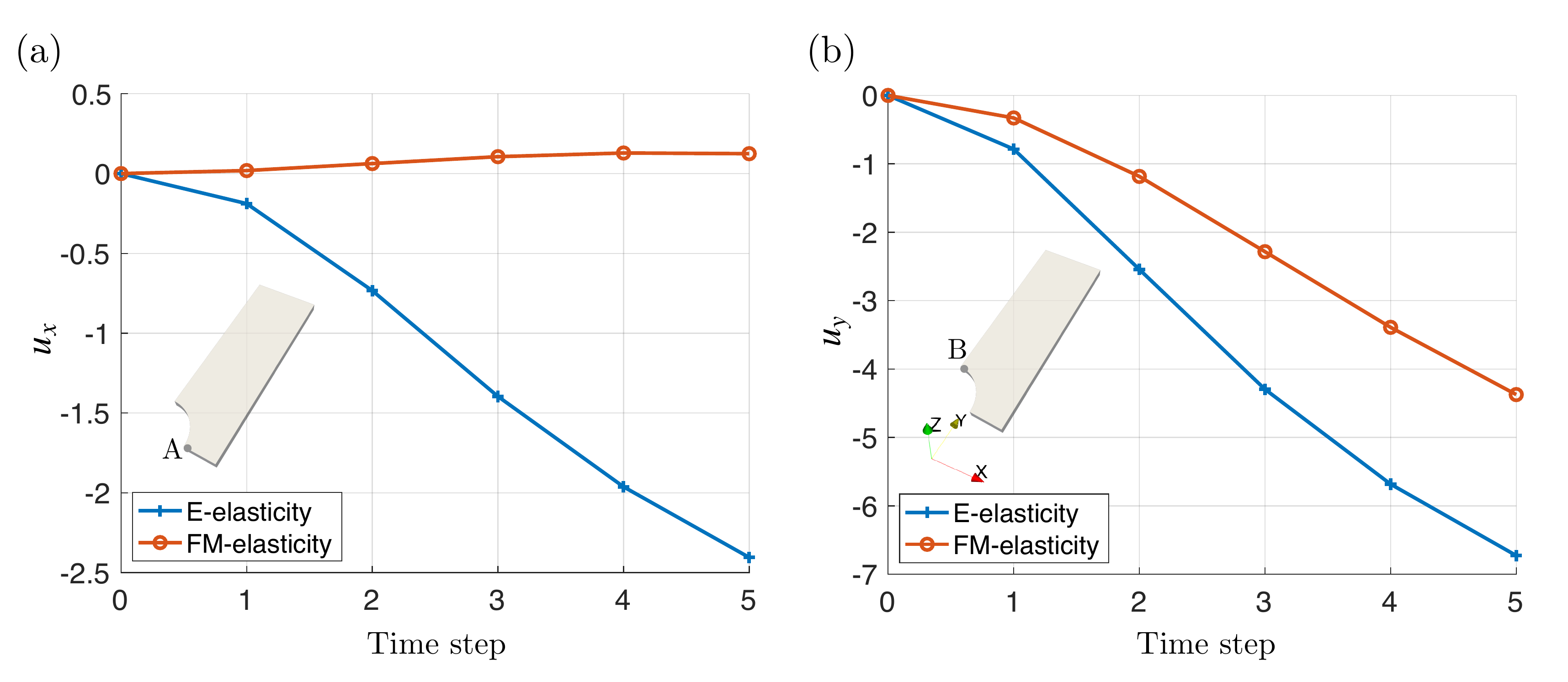}
    \caption{The (a) horizontal displacement of point A, and (b) the vertical displacement of point B for the strip with a hole problem over five time (load) steps.}
    \label{fig_strip_hole_e_fm_elasticity_plots}
  \end{figure}

The deformation induced by the applied potential difference is large, as shown in \fig{fig_strip_hole_e_fm_elasticity}.
The results for E-Elasticity in \fig{fig_strip_hole_e_fm_elasticity}(a) match those presented by \citet{Vu2007}.
The FM-Elasticity problem, shown in \fig{fig_strip_hole_e_fm_elasticity}(b), highlights important differences between the models.
The mechanical deformation in the vicinity of the hole differ significantly.
This is a consequence of the micromorphic response in FM-Elasticity that occurs due to the  inhomogeneous deformation field in this region, as shown in \sect{sec_strip_hole_m_elasticity} for the problem of M-Elasticity.
The differences between the theories are more clearly demonstrated in the plot of the horizontal and the vertical displacement of the points A and B, respectively, (see \fig{fig_strip_with_hole_geom_bcs}(b)) against the time (load) step shown in \fig{fig_strip_hole_e_fm_elasticity_plots}.

\subsection{Bending of a micro-cantilever beam}

\begin{figure}[htb!]
    \centering
    \includegraphics[width=0.6\textwidth]{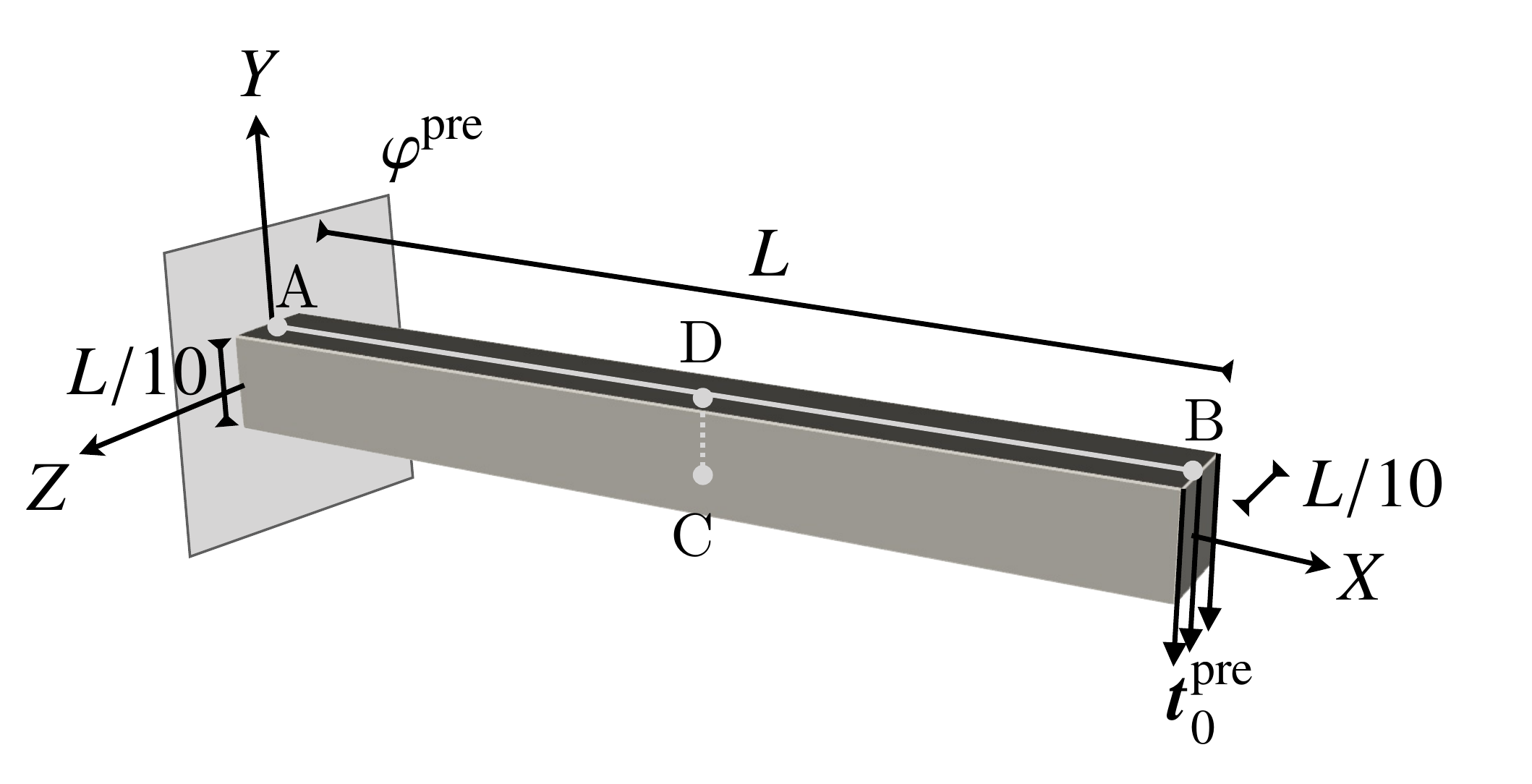}
    \caption{The geometry and boundary conditions for the micro-cantilever beam problem.}
    \label{fig_beam}
  \end{figure}

Consider the microscale cantilever beam of dimensions $L \times L/10 \times L/10$ shown in \fig{fig_beam}, where $L = \num{100}$.
The beam is fully fixed on the left face at $X=0$, that is, the macroscopic displacement $\gz{u} = \gz{0}$.
A traction $\gz{t}_0^\text{pre} = \num{-0.2}\gz{E}_2$ is applied to the right face at $X=L$ in a single load step.
The electric potential $\potential = \potential^\text{pre} = \num{0}$ at $X=0$.
The setup is typical of that used to quantify the flexoelectric response \citep[see e.g.][and the references therein]{Wenhui2001}.
The applied traction causes the cantilever to bend, thereby inducing a transverse strain gradient.
The polarization due to the flexoelectric effect is then determined from the measurement of the potential difference between two metallic plates on the upper and lower faces of the beam, as indicated by points C and D in \fig{fig_beam}.

  \subsubsection{M-Elasticity}

\begin{figure}[htb!]
    \centering
    \includegraphics[width=0.65\textwidth]{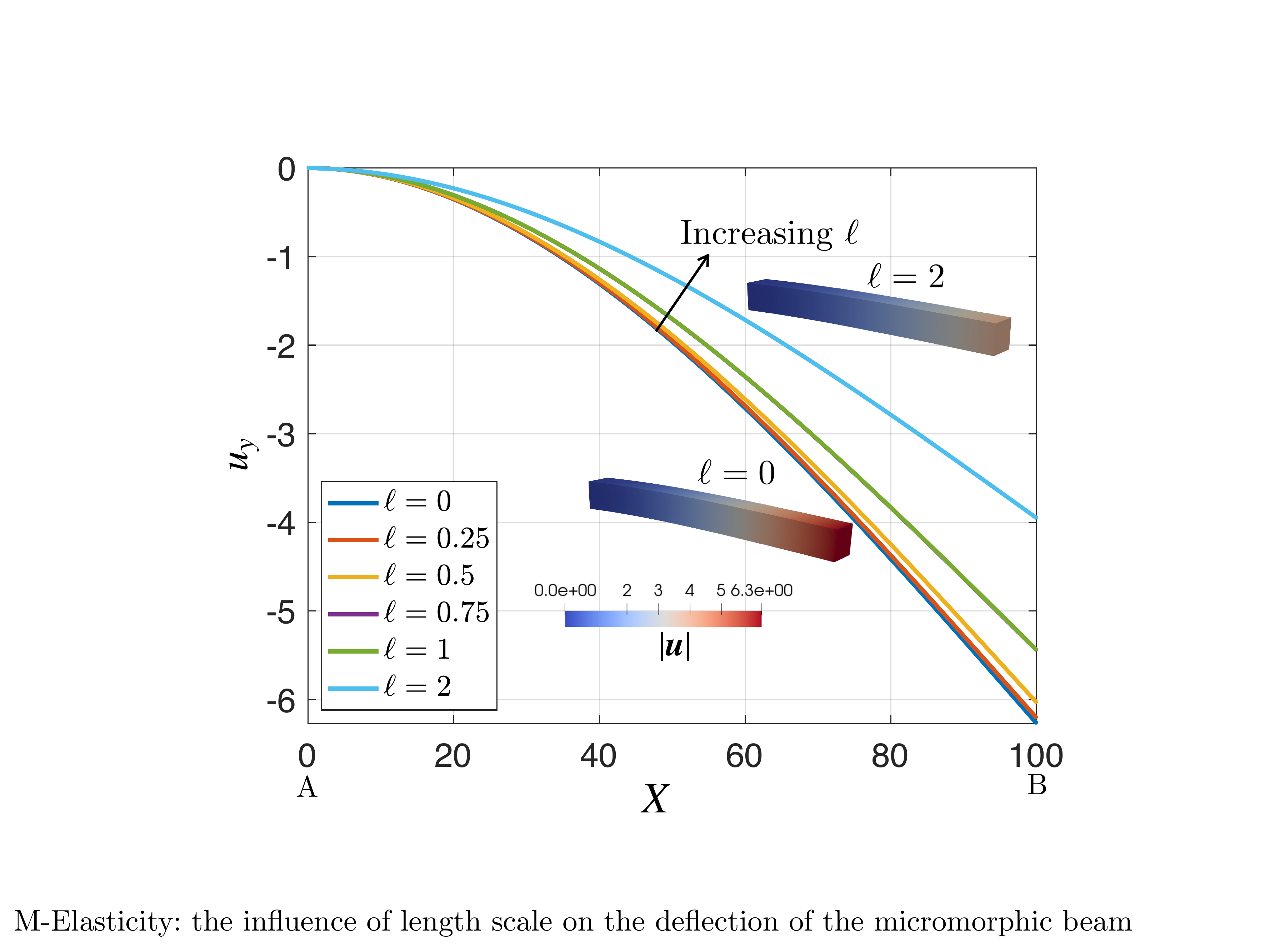}
    \caption{The vertical deflection $u_y$ along the line A--B for the cantilever beam for various choices of the length scale $\ell$.
    The problem is M-Elasticity.
    A plot of the deformed shape of the beam for the choices $\ell \equiv 0$ (Elasticity) and $\ell \equiv 2$ is also shown.
    The deformed shape is coloured by the magnitude of the displacement field.}
    \label{fig_beam_m_elasticity_plots}
  \end{figure}

The influence of the length scale $\ell$ on the vertical deflection of the beam $u_y$ due to the applied traction is investigated for the problem of M-Elasticity to determine the role of the micromorphic contribution in the absence of flexoelectric effects.
The vertical deflection along the line A--B for $\ell \in \{ 0; 0.25; 0.5; 0.75; 1; 2\}$ is shown in \fig{fig_beam_m_elasticity_plots}.
The choice of $\ell \equiv 0$ corresponds to (nonlinear) Elasticity.
An increasing length scale (decreasing specimen size) leads to a stiffer response, as expected.
As $\ell \to 0$ we recover the Elasticity response.

  \subsubsection{EM- and FM-Elasticity}

  The length scale is now fixed as $\ell \equiv 1$ (see \fig{fig_beam_m_elasticity_plots}) and the influence of the flexoelectric coefficient $\overline{\upsilon} \in \{0; 0.25; 0.5; 0.75; 1 \}$ on the response investigated.

\begin{figure}[htb!]
    \centering
    \includegraphics[width=\textwidth]{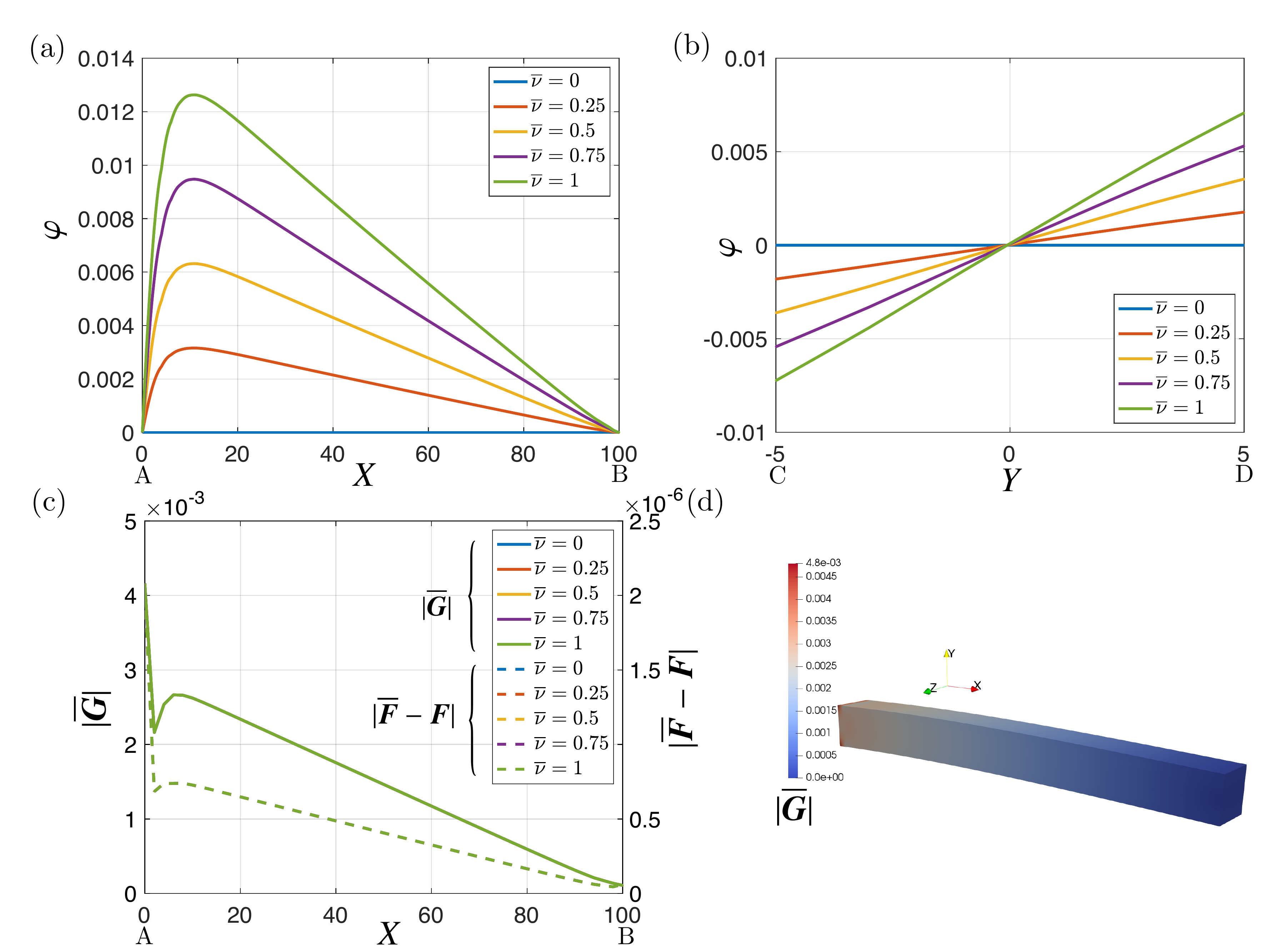}
    \caption{The distribution of the potential $\potential$ over (a) the horizontal line A--B and (b) the vertical line C--D, for various choices of $\overline{\upsilon}$.
    The distribution of the norm of the micro-gradient $\vert \mG \vert$ and the scale transition measure $\vert \mF - \F \vert$ along the line A--B is shown in (c).
    The distribution of the norm of the micro-gradient $\vert \mG \vert$ over the deformed cantilever beam is given in (d).}
    \label{fig_beam_fm_elasticity_plots}
  \end{figure}

  The distribution of the potential along the horizontal line A--B and the vertical line C--D (see \fig{fig_beam}) is shown in \fig{fig_beam_fm_elasticity_plots}(a) and (b), respectively.
  The choice of $\overline{\upsilon} \equiv 0$ corresponds to EM-Elasticity.
  For this case, and for the current choice of electric boundary conditions, the potential is zero in the beam.
  Choosing $\overline{\upsilon} > 0$ activates the flexoelectric effect.
  Increasing $\overline{\upsilon}$ linearly scales the magnitude of the distribution of the potential over the beam.
  This response can be understood from the distribution of the micro-gradient $\mG = \Grad \mF$ along the beam shown in \fig{fig_beam_fm_elasticity_plots}(c) and (d).
  The distribution of $\vert \mG \vert$ along the line A--B is essentially identical for all choices of $\overline{\upsilon}$.
  Thus the electric field has negligible influence on the micromorphic response in the current example.
  The choice of the energy associated with the flexoelectric effect in \eqn{psi_flexo_A} is linear in $\mG$.
  Hence the flexoelectric contribution to the dielectric displacement scales linearly with the flexoelectric coefficient $\overline{\upsilon}$, see \eqn{eD}.

 The micro-gradient $\mG$ exhibits a concentration at $X=0$ where the beam is macroscopically fully-constrained, see \fig{fig_beam_fm_elasticity_plots}(c) and (d).
 The macroscopic boundary condition results in a concentration in the macroscopic deformation field $\F$ and hence the micro-deformation $\mF$.

 Notice, however, the discrepancy between $\vert \mF - \F \vert$ at the boundary shown in \fig{fig_beam_fm_elasticity_plots}(c).
 Recall that the scale bridging energy $\psi_0^\text{scale}$ in \eqn{psi_scale} contains the term $[\mF - \F]$.
 The micro-deformation $\mF$ is a field variable (a nodal quantity in the finite element description) while the deformation gradient $\F$ is computed from the displacement field and evaluated at the quadrature points of the finite element mesh.
 This mismatch leads to the inability to tie $\mF$ to $\F$ more closely in the presence of a concentration in the macroscopic deformation field irrespective as to the choice of the penalty term $p$.
 Several choices for $p$ were investigated and all produced similar behaviour.

\section{Discussion and conclusion} \label{sec_discuss_conclusion}

A novel micromorphic formulation for flexoelectricity has been presented.
The formulation has been implemented within a conventional $C^0$-continuous finite element setting.
The Dirichlet principle has been applied to reveal the structure of the governing relations and the boundary conditions.
The formulation allows for a spectrum of different model problems to be considered by the appropriate restriction of the constitutive parameters.
Details of the finite element approximation have been given.
The theory has been elucidated via a series of numerical example problems.
The cantilever beam example demonstrated the complex interaction between the mechanical size effect and the flexoelectric response.
%Continuum modelling is therefore crucial to provide further insight into experimental observation.

The influence of the free space surrounding the solid material has been ignored, as is often the case for piezoelectric materials \citep[see e.g.][]{Poya2015}.
This is not however the case for electro-active polymers where the free space contribution can be significant \citep{Vogel2014}, and was accounted for by \citet{Yvonnet2017} in their model of flexoelectricity.
Therefore, the framework presented will be extended to consider the free space.
The approach will follow our previous work on E-Elasticity \citep{Pelteret2016}.

The current work presented a mathematical and numerical model for flexoelectricity.
The numerical model has been validated for E-Elasticity and M-Elasticity using benchmark problems in the literature.
The validation of the FM-Elasticity model against experiment is critical and will be considered in future work.
This will allow one to decide on the correct form of the flexoelectric energy and the choice of the relevant constitutive parameters.
Experimentally measured uncertainty in the geometry and the constitutive parameters of the fabricated components should be accounted for in the model.

The micromorphic framework presented could readily be extended to describe the converse flexoelectric effect via the introduction of a micromorphic electrical field $\overline \Efield$ and its gradient $\overline{\mathbb{G}} = \Grad \overline \Efield$.
The scale-bridging energy for the converse effect would involve the norm $\vert \overline \Efield - \Efield \vert$ and take a form similar to \eqn{psi_scale}.
The form of the energy describing the converse flexoelectric effect would include $\overline{\mathbb{G}}$ and some measure of the macroscopic deformation.

The choice of the optimal functional setting for the problem of flexoelectricity remains an open challenge.
A careful mathematical and computational study will provide further insight and is recommended.
This may also reveal an approach to better control the scale transition parameter  in the vicinity of concentrations in the macroscopic fields.

\section*{Acknowledgements}
PS and AM gratefully acknowledge the support provided by the EPSRC Strategic Support Package: Engineering of Active Materials by Multiscale/Multiphysics Computational Mechanics - EP/R008531/1.
DD was partly supported by the German Research Foundation (Deutsche Forschungsgemeinschaft, DFG), grant DA 1664/2-1.

\appendix

\section{Form of the kinetic measures}\label{appendix_linearisation}

The form of the kinetic measures introduced in Table \ref{table_kinetics} and entering the residual equations \eqref{R_motion} -- \eqref{R_potential} are now given for the choice of constitutive relations made in \sect{sec_const_relations}.
Prior to this, some useful relations are recalled.

\subsection{Useful relations}

\begin{align*}
    \dfrac{\partial \left[ \F : \F \right ]}{\partial \F} = 2 \F \, ,
        &&
    \dfrac{\partial \left[ J \right] }{\partial \F} = J \invF\trns \, ,
    \\
    \dfrac{\partial \left[ \ln J \right] }{\partial \F} = \invF\trns \, ,
        &&
    \dfrac{\partial \left[ \ln^2 J \right] }{\partial \F} = 2 \ln J \invF\trns  \, ,
    \\
    \dfrac{\partial \left[ \gz C \right] }{\partial \F}
        = \gz I \underline{\otimes} \F\trns + \gz F \trns \overline{\otimes} \gz I \, ,
        &&
    \dfrac{\partial \invF \trns }{\partial \F}
        =
        -\invF\trns \underline{\otimes} \invF \, ,
    \\
    \dfrac{\partial \gz B}{\partial \gz C}
        =
        -\dfrac{1}{2} \left[ \gz B \underline{\otimes} \gz B + \gz B \overline{\otimes} \gz B  \right] \, ,
        &&
    \dfrac{\partial \gz B}{\partial \gz F}
        =
        -\left[
            \invF \overline{\otimes} \gz B
            +
            \gz B\underline{\otimes} \invF
        \right] \, .
    % \dfrac{\partial \gz C  : [\Efield \otimes \Efield]}{\partial  \F}
    %     & =
    %     2 \left[ \Efield \cdot \F\trns \right] \otimes \Efield
    % &
    % \dfrac{\partial J \gz B  : [\Efield \otimes \Efield]}{\partial  \F}
    % & =
    % 2 \left[ \Efield \cdot \F\trns \right] \otimes \Efield
    % \dfrac{\partial \gz B  : [\Efield \otimes \Efield]}{\partial  \F}
    %     & =
    %     2 \left[ \Efield \cdot \F\trns \right] \otimes \Efield \\
\end{align*}

\subsection{Kinetic measures}

\subsubsection*{Macroscopic Piola stress $\Ptot$}

The macroscopic Piola stress $\Ptot$ is given by
\begin{align*}
    \Ptot
    &= \D_{\F} U_0
    = \D_{\F} \psi_0^{\text{elast}}
        + \D_{\F} \psi_0^{\text{flexo}}
        + D_{\F} E_0
    = \D_{\F}\psi_0^\text{mac}
        + \D_{\F} \psi_0^\text{scale}
        + \D_{\F} \psi_0^{\text{flexo}}
        + D_{\F} E_0
    \intertext{where}
    D_{\F}\psi_0^\text{mac}
        &=
    \left[ \left[ \lambda \ln J - \mu \right] \invF \trns
        + \mu \F \right]
        + 2 \epsilon_0 \beta  \gz C \cdot \Efield \, ,
    \\
    \D_{\F} \psi_0^\text{scale}
        &=
        - p \left[ \mF - \F \right] \, ,
    \\
    \D_{F_{iJ}} \psi_0^{\text{flexo}}
        &=
        -\eta
        \biggl[
        \Efield_{M}
        \left[
            \invF\trns \underline{\otimes} \invF
        \right]_{mMiJ}
        \left[
            \mG:\gz B
        \right]_{m}
        +
        \left[ \left[ \gz{f}\trns \cdot \Efield  \right] \cdot \mG\right]_{NO}
            \left[
                \invF \overline{\otimes} \gz B
                +
                \gz B\underline{\otimes} \invF
            \right]_{NOiJ}
        \biggr]
        \gz e_i \otimes \gz E_J \, ,
    \\
    D_{\F}E_0
    & =
    -\dfrac{1}{2} \eta \epsilon_0 J
    \left[
        [\gz{B} : \Efield \otimes \Efield] \left[ \invF \trns \right]_{iJ}
        -
        \left[
            \invF \overline{\otimes} \gz B
            +
            \gz B\underline{\otimes} \invF
        \right]_{MNiJ} \left[ \Efield \otimes \Efield \right]_{MN}
    \right] \gz e_i \otimes \gz E_J \, .
\end{align*}

\subsubsection*{Micromorphic Piola stress}

The micromorphic Piola stress $\mP$ is given by
\begin{align*}
    \mP & =  \D_{\mF} U_0
        = \D_{\mF} \psi_0^{\text{elast}}
        =  \D_{\mF} \psi_0^\text{scale}
    \\
    &   = p \left[ \mF - \F \right] \, .
\end{align*}

\subsubsection*{Micromorphic double stress}

The micromorphic double stress $\mQ$ is given by
\begin{align*}
    \mQ & =  \D_{\mG} U_0
        = \D_{\mG} \psi_0^{\text{elast}}
        + \D_{\mG} \psi_0^\text{flexo}
    \\
    &= \mu \ell^2 \mG + \nu \left[ \invF  \trns \cdot \Efield \right] \otimes \gz{B} \, .
\end{align*}

\subsubsection*{Dielectric displacement}

The dielectric displacement $\eD$ is given by
\begin{align}
    \eD = - \D_{\Efield} U_0
        &=
        -2 \epsilon_0 \left[ \alpha  \gz I  + \beta  \gz C - \dfrac{1}{2} \eta \epsilon_0 J  \, \gz{B}\right ]
            \cdot  \Efield
        - \upsilon \gz{f}  \cdot \mG : \gz{I}
            \label{eD} \, .
\end{align}

% \clearpage

\bibliographystyle{unsrtnat}
\bibliography{bibliography}

\end{document}